\documentclass[fleqn,usenatbib]{mnras}
\usepackage{graphicx}
\usepackage{xcolor}
\usepackage{hyperref}
\usepackage{amsmath, amssymb}
\usepackage{natbib}
\usepackage{placeins}
\usepackage{booktabs}
\usepackage{units}
\usepackage[caption=false]{subfig}
\usepackage{float}
\usepackage[normalem]{ulem}

\newcommand{\thetaobs}{\ensuremath{\theta_{\text{obs}}}}
\newcommand{\mej}{\ensuremath{M_{\text{ej}}}}
\newcommand{\xlan}{\ensuremath{X_{\text{lan}}}}
\newcommand{\vej}{\ensuremath{v_{\text{ej}}}}

\title[El-CID]{
    El-CID: A filter for Gravitational-wave Electromagnetic Counterpart
    Identification
}
\author[D. Chatterjee et al.]{
Deep Chatterjee,$^{1,2}$\thanks{E-mail: deep1018@illinois.edu}
Gautham Narayan,$^{1,2}$
Patrick D. Aleo,$^{1,2}$
Konstantin Malanchev,$^{2,3}$
\newauthor~and Daniel Muthukrishna$^{4}$
\\
$^{1}$Center for AstroPhysical Surveys, National Center for Supercomputing Applications, Urbana, IL, 61801, USA\\
$^{2}$Department of Astronomy, University of Illinois at Urbana-Champaign, 1002 West Green Street, Urbana, IL 61801, USA\\
$^{3}$Lomonosov Moscow State University, Sternberg Astronomical Institute, Universitetsky pr.~13, Moscow, 119234, Russia\\
$^4$Institute of Astronomy, University of Cambridge, Madingley Road, Cambridge CB3 0HA, UK
}

\begin{document}
\label{firstpage}
\pagerange{\pageref{firstpage}--\pageref{lastpage}}
\maketitle

\begin{abstract}
As gravitational-wave interferometers become more sensitive and probe ever more distant reaches, the number of detected binary neutron star mergers will increase. However, detecting more events farther away with gravitational waves does not guarantee corresponding increase in the number of electromagnetic counterparts of these events. Current and upcoming wide-field surveys that participate in GW follow-up operations will have to contend with distinguishing the kilonova from the ever increasing number of transients they detect, many of which will be consistent with the GW sky-localization. We have developed a novel tool based on a temporal convolutional neural network architecture, trained on sparse early-time photometry and contextual information for Electromagnetic Counterpart Identification (El-CID). The overarching goal for El-CID is to slice through list of new transient candidates that are consistent with the GW sky localization, and determine which sources are consistent with kilonovae, allowing limited target-of-opportunity resources to be used judiciously. In addition to verifying the performance of our algorithm on an extensive testing sample, we validate it on AT2017gfo - the only EM counterpart of a binary neutron star merger discovered to date - and AT2019npv - a supernova that was initially suspected as a counterpart of the gravitational-wave event, GW190814, but was later ruled out after further analysis.
\end{abstract}


\section{Introduction}
While the discovery of gravitational waves (GWs) has been rather recent,~\citep{Abbott_2016} the field has moved swiftly since. The discovery rate has increased more than an order of magnitude over the last five years, from three events in the first observing run to more than thirty in the most recent catalog~\citep{2020arXiv201014527A}. With the start of the third LIGO/Virgo Collaboration (LVC) observing run, O3, in 2019 GW astrophysics has moved into the public alert era. LVC has created critical   infrastructure \footnote{\url{https://emfollow.docs.ligo.org/userguide/}} to broadcast discovery information in near real-time so that GW candidates can be followed-up in the electromagnetic (EM) spectrum. Identifying the EM counterpart enables much richer multi-wavelength studies of these events, providing much stronger constraints on the evolution of the progenitor system during the kilonova explosion, and the resulting ejecta composition. This, in turn, allows us to understand the role of these events in the broader context of galaxy evolution. Combining EM and GW data may also improve the utility of these events as standard sirens,~\citep{schutz_1986,holz_hughes_2005} helping to break the degeneracy between their distance and inclination, allowing their use as cosmological probes in the nearby Universe~\citep{2017Natur.551...85A}.

The resources expended by the community on multi-messenger astrophysics is warranted. The first ever observed binary neutron star (BNS) merger, GW170817,~\citep{gw170817} discovered on August 17, 2017 was followed by a short gamma-ray burst, GRB~170817A,~\citep{2017ApJ...848L..14G} and the first joint observation of the \emph{kilonova} (KN), AT~2017gfo across the electromagnetic spectrum\footnote{There have been reports of potential kilonovae earlier from GRBs, see for example \cite{Tanvir_2013}}\citep{mma_2017, arcavi_2017, Coulter_2017, kasliwal_2017, Lipunov_2017, Soares_Santos_2017, Tanvir_2017, Valenti_2017}. Subsequent observations of this one event over the next hours through months, by facilities worldwide revealed that it was the site of heavy element nucleosynthesis. It provided the first observational evidence for the several decade-old hypothesis that compact object mergers can be the progenitors of short GRBs and KNe~\citep{lattimer_1974,1976ApJ...210..549L,1984SvAL...10..177B, 1998ApJ...507L..59L,2010MNRAS.406.2650M}.The wealth of data of GW170817 then reminds us of the promise of the science that is enabled by identifying the EM counterparts of these events. The third observing run reported a total of 56 candidates released publicly. The GWTC-2 catalog~\citep{2020arXiv201014527A} has reported more events than those discovered during real-time analysis. However, despite many advances and significant effort by the community, the O3 run concluded without another coincident detection of an EM counterpart. One reason is the larger distances at which the promising O3 candidates were reported or the fact that they had large sky-localization areas (see \cite{2020NatAs...4..550C}, for a summary).

% Some of the rich science obtained from this single event included the origin of heavy
% elements beyond iron, a \emph{standard siren} measurement of the Hubble constant,~\citep{2017Natur.551...85A}
% the constraints on the neutron star equation of state,~\citep{Abbott_2018},
% constriants on the post-merger ejecta~\citep{Abbott_2017} to name a few.

The next few years brings a growing ground-based GW detector network
that is reaching their design-sensitivity. There are $\mathcal{O}(10)$ BNS
mergers expected in the fourth observing run~\citep{2018LRR....21....3A} with a average range $\sim 190$~Mpc. In the design sensitivity era this is expected to increase to $\sim 330$~Mpc, which potentially implies a further $\sim 5$-fold increase in number of events detected. This increase in the number of events from GW facilities is matched by complementary advances in the optical and near-infrared, enabled by next generation telescope facilities like the Vera Rubin Observatory~\citep{Ivezi_2019}~\footnote{The Vera Rubin Observatory was previously known as Large Synoptic Survey Telescope (LSST). Currently, LSST refers to the 10-year Legacy Survey of Space and Time that the observatory will conduct.}
and the \emph{Nancy Grace Roman Space Telescope}~\citep[\emph{NGRST},][]{Kasdin_2020}. Crucially for multi-messenger astrophysics, because these new facilities not only go deeper and cover a larger area on the sky at higher cadence than existing surveys, they have also catalyzed upgrades to existing facilities, making them more responsive. For example, telescope and observatory management systems \citep[TOMS,][]{TOMS} allow programmatic access to queue schedule a large array of facilities such as the Astrophysical Events Observatory Network \citep[AEON,][]{AEON}, increasing the likelihood of obtaining additional photometry and spectroscopy for GW sources within a few hours of explosions.

However, \emph{identifying} the electromagnetic counterparts of BNS mergers in O4 and beyond presents a greater challenge to time-domain astronomy than run O3 since kilonovae evolve much faster than the typical cadence of these surveys. Discovering the counterpart requires that we are able to distinguish it from all of the other transient events also detected contemporaneously. The astronomical community has already wrestled with this challenge in O3 when several initially ``interesting'' candidates that were ruled out later, demonstrating the needle in a haystack nature of the problem. As a result, a targeted search for exotic transients like KNe will be polluted by contaminant ``vanilla'' objects, like supernovae. Followup time on large aperture facilities for these faint objects is extremely limited and it will not be possible to whittle down the list of potential counterparts by filtering out these contaminants using spectroscopy. While Integral Field Unit (IFU) spectroscopy is more promising in this respect, these facilities generally do not have large fields of view. Simply put, if EM followup in O3 was hard, followup in O4 is only going to get harder. Therefore, it is crucial to develop tools that will filter interesting transients from the initial sparse photometry itself. This is the primary motivation for this work.

Given the large number of candidate counterparts from EM surveys, and the need for fast followup to monitor the rapid evolution of kilonovae, we focus on machine learning techniques to identify the counterpart photometrically. Several attempts at automated photometric classification of objects like supernovae have been reported earlier. Initial efforts relied on template  fitting~\citep{kessler2010supernova,2010ApJ...717...40K}. Subsequent approaches utilized features extracted from such fits as features to train machine-learning (ML) algorithms~\citep{2018ApJS..236....9N}. \citet{2016ApJS..225...31L} utilized a non-parametric wavelet decomposition-based approach. The Photometric LSST Astronomical Time Series Classification Challenge (PLAsTiCC)~\citep{plasticc,plasticcresults} was a global effort involving competing teams to classify $\mathcal{O}$(millions) of simulated transients and variable lightcurves using machine learning (ML). All these attempts, however, only considered classification using the complete phase information of the lightcurve i.e. what the type of the object is, given that we have seen it rise, reach maximum light and fade away. These approaches are suitable to produce a complete catalog of photometrically-identified kilonovae retrospectively, and while this is important for studying the population of these objects, it does not enable real-time discovery and followup.

Photometric classification performance using only sparse initial photometry is considerably worse than performance when using the complete phase information. While early-time photometric information is necessarily sparse, and likely to be low S/N, wide-field optical and infrared surveys likely have access to considerable \emph{contextual} information at the location of candidates. Classification of \emph{transients} using only contextual information from their host galaxies, such as colors, light profile and redshift was first demonstrated by \cite{2013ApJ...778..167F}. Recently, \cite{Gagliano_2021} further improved this approach considering an order of magnitude larger sample of the spectroscopically classified galaxies. Combining both early-time information and contextual information may provide a method to tackle the needle-in-a-haystack problem.

Of particular relevance to our work is the RAPID early-time classifier described in \cite{rapid_2019} (hereafter MU19). RAPID employs deep neural-networks (DNN) considering the early-time data as a time-series and mapping it to the transient class. The advantage is that the classification is provided with every epoch of data acquired. The computational cost is only encountered during training the DNN, with real time classification during nightly operations. The classifier can be further guided by providing contextual information. For example, in case of supernova classification, the knowledge about the redshift guides the classifier to give a better performance. Here, we leverage this feature -- while the redshift may not be available for new objects from telescopes during the time of discovery of GW, there are several other data products that are provided during a GW discovery like the GW sky localization. Features of the skymap and its correlations with the objects discovered during follow-up, can serve a similar role and help in the filtering out the impurities which can save precious telescope resources. In this study, we modify the architecture used in RAPID, tuning it as a binary classifier for Electromagnetic Counterpart identification (El-CID) i.e., to filter KNe from other transients during follow-up of GW candidates.

The organization of the paper is as follows. In Sec.~\ref{sec:snana} we describe the generation of our training data. We create an end-to-end training set starting with simulating BNS mergers in GWs. We generate skymaps for mergers that are detected by LIGO/Virgo. We then simulate KNe lightcurves associating the BNS mergers to KNe. We also generate other explosive transients which are supposed to be the contaminants during follow-up. In Sec.~\ref{sec:training} we describe the training features and the contextual information that we provide during training. In Sec.~\ref{sec:neural-net}, we talk about the neural network architecture and changes compared to RAPID. In Sec.~\ref{sec:results}, we evaluate the performance of the classifier on
both simulated and real data. We find expected classification for real events like AT~2017gfo, counterpart of GW170817, and AT~2019npv, a supernova that was in the field of view of GW190814 and was initally considered of interest.
We conclude in Sec.~\ref{sec:conclusion} highlighting the use case of a tool like El-CID for future follow-up operations.

\section{Simulating transients}\label{sec:snana}
\begin{figure}
    \centering
    \includegraphics[width=1.0\columnwidth]{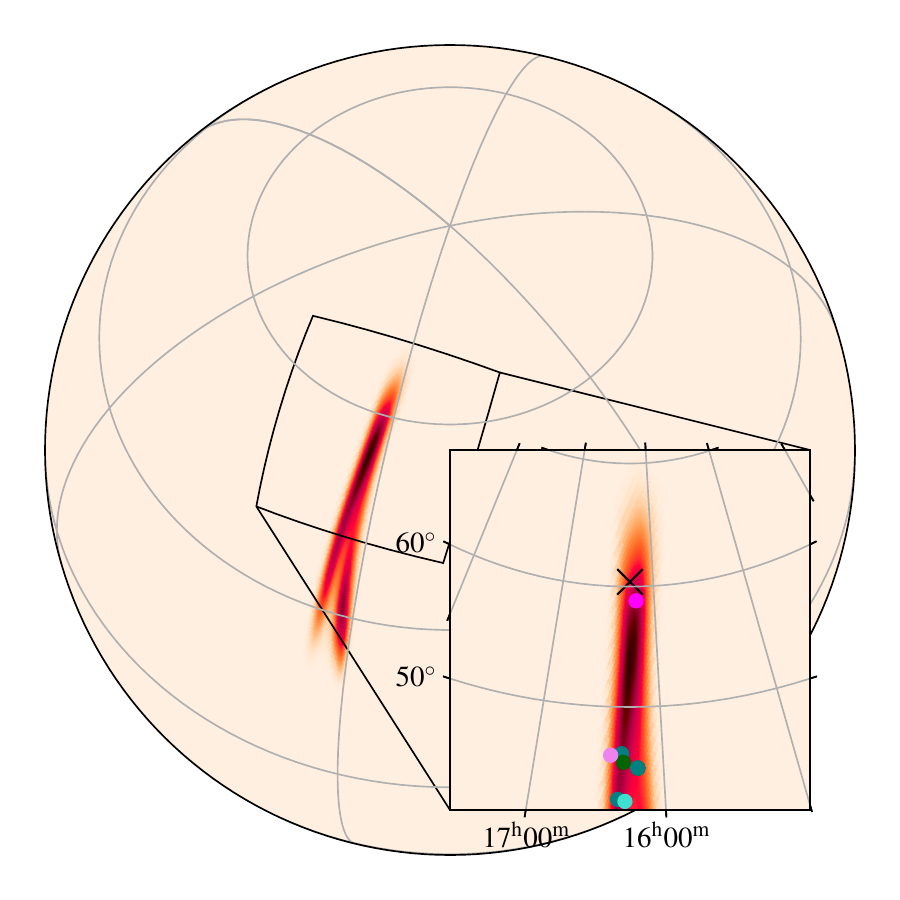}
    \caption{An example scenario: the GW skymap associated with the KN
    is shown in the globe projection. The inset shows the sky locations
    of simulated transients. The KN is shown in the `x' symbol, the other
    objects are shown using circles.
    }
    \label{fig:mock_skymap}
\end{figure}
\begin{figure}
    \centering
    \includegraphics[width=1.1\columnwidth, trim=0cm 1cm 0cm 1cm, clip]
        {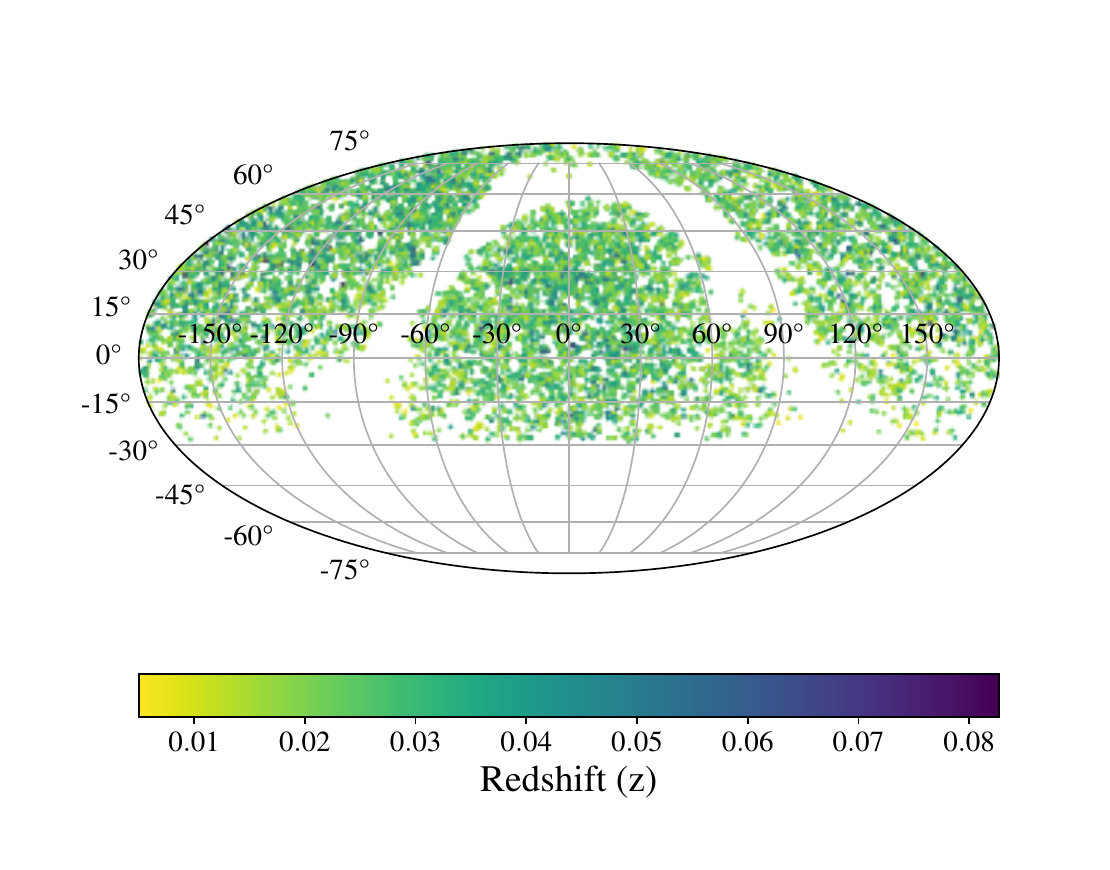} \\
    \includegraphics[width=1.1\columnwidth, trim=0cm 1cm 0cm 1cm, clip]
        {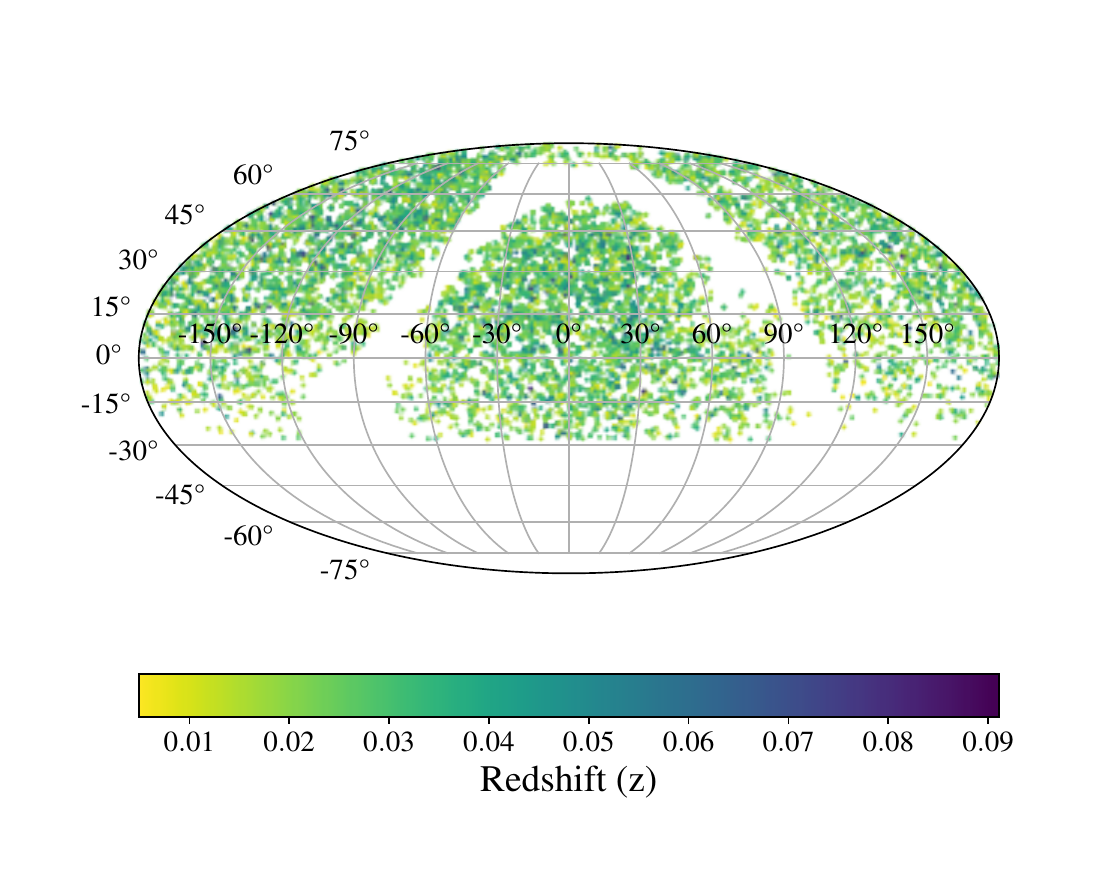} \\
    \includegraphics[width=1.1\columnwidth, trim=0cm 0cm 0cm 0.5cm, clip]
        {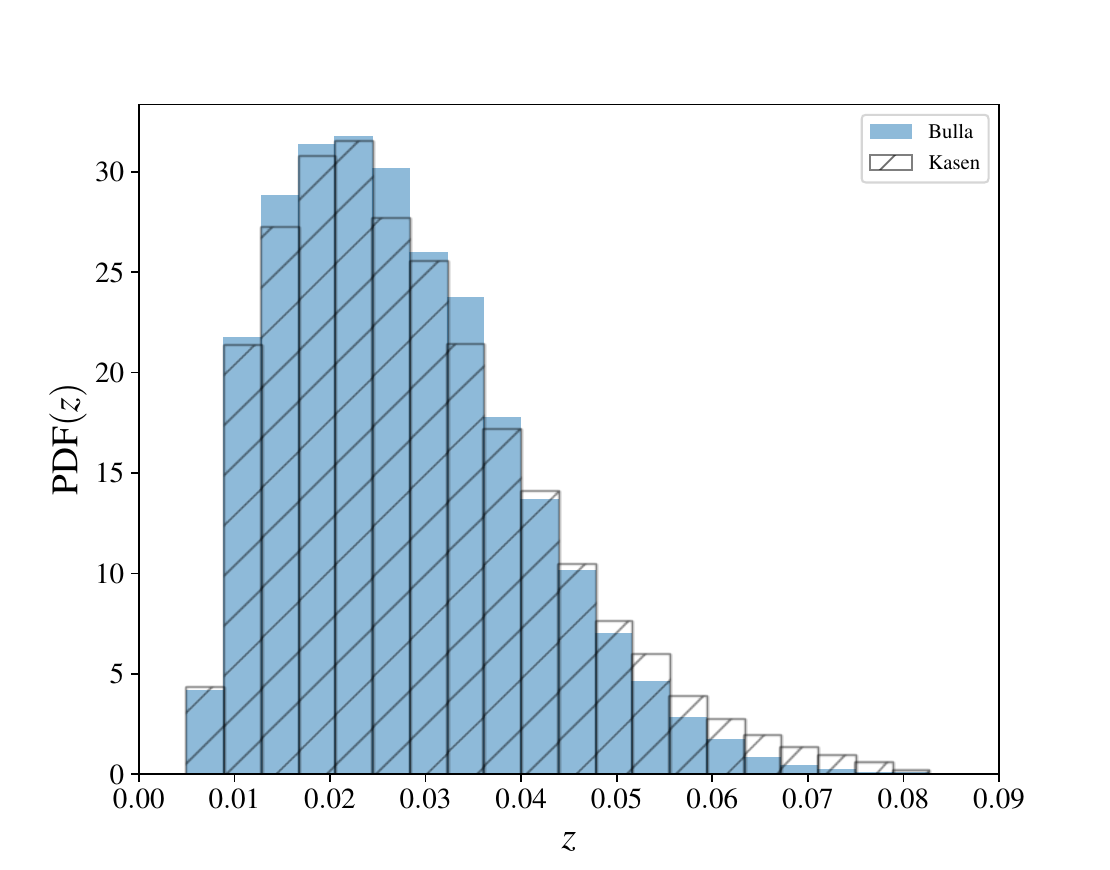}
    \caption{\textbf{Upper panel}: Distribution of the recovered Bulla KNe in
    the sky in Mollweide projection. The shading is based on the simulated redshift
    of the source. \textbf{Middle panel}: Same as above for the Kasen models.
    \textbf{Lower panel}: Distribution of the redshift for the recovered KNe for
    both models. The Bulla model case is shown in the filled histogram, while that
    for the Kasen model is shown in the hatched histogram. We observe that there is
    no preference towards detecting a specific model.}
    \label{fig:kasen_bulla_distribution}
\end{figure}

The situation we are considering is shown visually in 
Fig.~\ref{fig:mock_skymap} using a simulated GW sky-localization (skymap). The shading represents
the probability density in the sky coordinates. If this is from
a binary neutron star (BNS) merger, the associated KNe will likely lie in a region of high probability. We show the true location of the source with the $\times$ symbol. However, when attempting to identify the counterpart, we must contend with other unrelated transients that are temporally coincident and near the same high probability region defined in the GW alert. These events will also be discovered by surveys during their operations, and are shown by the solid circles. All these objects will have some photometric data; those with data before the GW trigger time are disqualified from being associated with the GW trigger. At the same time, if the first observation is made $\gtrsim 1$ week after the GW trigger, it is also likely unrelated since most KNe are expected to peak in optical bands $\sim 1-3$ days. Coincidental transients that satisfy the above criteria are the contaminants from which we must distinguish the kilonova. 

During the initial hours through a few days after the GW alert, there will be a paucity of imaging data for all these sources for robust photometric classification. However, there is helpful contextual information that can
guide a classification algorithm with the sparse data from the initial epochs of observation. Provided the LVC analysis pipeline is not biased, in general we can expect the true counterpart to be closer to the high-confidence region of the skymap, and the line-of-sight probability to the true event to be slightly higher than in random directions that are scattered about the true event. Given the location of the candidates, we can use the line-of-sight probability based on the skymap, and the angular offset from
the mode of the skymap -- as important contextual information to supplement the sparse initial photometry. While neither exclusively can provide a robust classification, in Sec.~\ref{sec:results}, we demonstrate that combining both using the neural network architecture in Sec.~\ref{sec:neural-net} does result in a reliable classification for the simulations we describe below, as well as GW170817 data.

To create an end-to-end realistic scenario like the one mentioned above, we generate a exhaustive set of simulated transients. This includes KNe and other explosive transients. We use the Supernova Analysis software package, SNANA~\citep{snana} for this purpose. While the scope of SNANA is broad, ranging from fitting lightcurves, classifying SNe, tools for SNe cosmology, and simulating explosive and variable transients, it is the last role that we are concerned with here. Given an empirical or spectral energy density (SED) model of an explosive transient along with the details of observing strategy and conditions of a survey i.e., its location, cadence, instrument thresholds, SNANA simulates lightcurves that would be observed by the survey including measurement errors. SNANA has been widely used in the literature for such simulations; a notable effort has been in generating $\sim 10^6$ of simulated transients and variable lightcurves for the Photometric LSST Astronomical Time Series Classification Challenge (PLAsTiCC)~\citep{plasticc}. Other efforts regarding tools to simulate lightcurves include {\tt{simsurvey}}~\citep{Feindt_2019}. Also, see~\cite{Chatterjee_2019} for a more rigorous treatment of detection efficiency of related to supernova rates. 

\subsection{Simulating KNe}
\begin{figure}
    \centering
    \includegraphics[width=1.0\columnwidth]{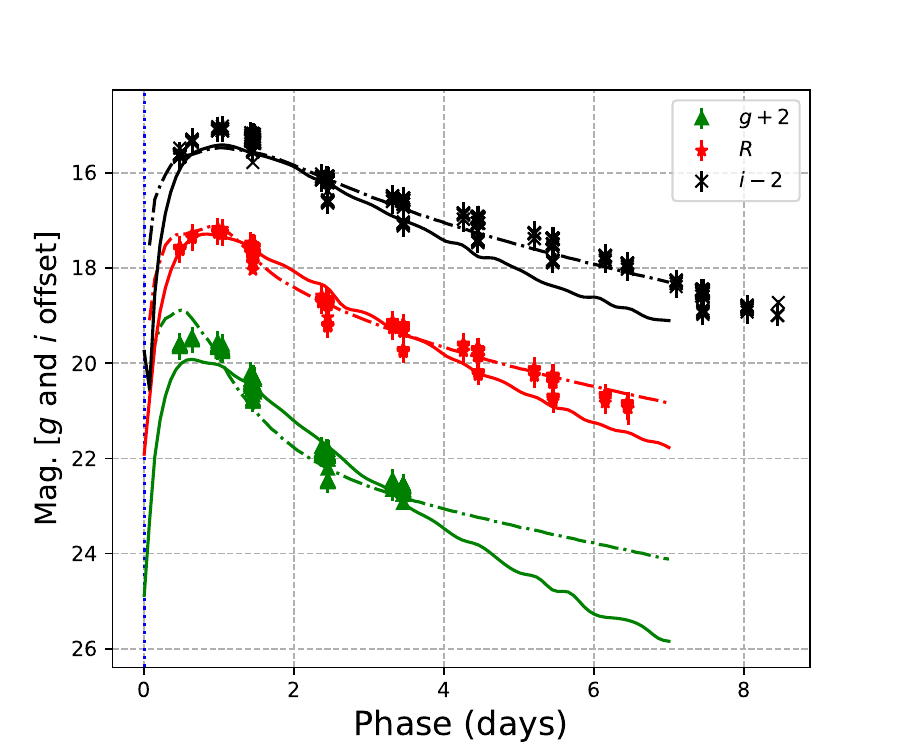}
    \caption{BU19 (solid curves) and KN17 (dotdash curves) models overplotted along with
    the DECam data from AT~2017gfo (offset for clarity). For the models in this figure, We use the best fit
    parameter model of the BU19 reported in~\citet{coughlin_2020}. For the KN17 model, we use
    one available publicly at \url{https://github.com/dnkasen/Kasen_Kilonova_Models_2017};
    \texttt{knova\_d1\_n50\_m0.025\_vk0.25\_fd0.3\_Xlan1e-6.0\_vs0.32\_ns12.0.h5}
    }
    \label{fig:gw170817_lightcurve_fits}
\end{figure}
For our KNe, we use the spectral energy density
(SED) model calculated on a kilonova model grid using the radiation transport code 
POSSIS~\citep[hereafter BU19]{bulla_2019}, used earlier in~\cite{coughlin_2020}. We have incorporated this
SED model grid in SNANA, and it can be accessed publicly as a part of the SNANA package data.\footnote{
\url{https://zenodo.org/record/4015340}} Each representative SED model is
parameterized by the ejecta mass from the merger, $\mej$, the fraction of the lanthanide-rich
component characterized by its half opening angle, $\Phi$,
and the observing angle with respect to the line of sight, $\thetaobs$ ($\cos\thetaobs=0$ being
edge-on, while $\cos\thetaobs=1$ being face-on). We also consider the KNe SED models by \citet{Kasen_2017} (hereafter KN17), in which case the SED is parameterized by the ejecta mass, $\mej$, the lanthanide fraction, $\xlan$, and the ejecta velocity, $\vej$. There is no observing angle dependence for the KN17 models. Given that there remains only one KNe definitively observed at electromagnetic wavelengths, there remain considerable model uncertainties, ultimately corresponding to different descriptions for the same physical system. In Fig.~\ref{fig:gw170817_lightcurve_fits}
we show example lightcurves from KN17 and BU19 models comparing them to AT~2017gfo data from DECam.~\footnote{
Produced using some \texttt{sncosmo} wrapper code available at: \url{https://github.com/deepchatterjeeligo/bulla-models}.}
We simulate an equal number of events ($\sim 6\times 10^6$ each) using both KN17 and BU19 models, so as to not bias our machine learning algorithm towards either model. As the sample of observed kilonovae grows, we expect these different parametrizations to converge, and we can update the models as needed. 
We distribute the simulated events uniformly across the sky, and volume assuming a flat $\Lambda$CDM cosmology with a Hubble constant, $H_0 = 70\;\text{km s}^{-1}/\text{Mpc}$, and matter density, $\Omega_m = 0.3$. We consider an observing strategy based on ZTF's public ``MSIP survey'' observations included in the third data release of ZTF (ZTF DR3).\footnote{ZTF data releases includes observations from both ``public'' and ``private'' surveys. The public survey was the source of alerts sent to brokers and was used to build a survey with predictable average cadence.} 
While ZTF has had subsequent data releases, we are only concerned with those that overlap with the LVC O3 run, so that our simulations reflect the published sensitives of both GW and EM surveys. 

We simulate the sky brightness using magnitude limit data from ZTF DR3 and point spread function size distribution from Fig.~6 of \citet{2019PASP..131a8002B}. We evaluate the detection efficiency as a function of signal-to-noise using the distribution of the measured magnitude errors from the ZTF DR3 data release (see Appendix~\ref{appendix:efficiency_ztf} for details). SNANA uses this information to determine the fraction of objects recovered at every epoch of detection. This process therefore includes realistic weather losses at the Palomar site. The distribution of the detected KNe in the sky is shown in Fig.~\ref{fig:kasen_bulla_distribution} where the upper (middle) panel correspond to the BU19 (KN17) models. We find that the detectability peaks around a redshift of $\sim 0.025$,
corresponding to a luminosity distance of $\sim 110$ Mpc. 
% GN - below is useful info - we should include it.
% This is lower than the angle
% averaged BNS merger range for GW which is expected $\sim 190$ Mpc in O4. With the Virgo
% detector sensitivity also $\sim 120$ Mpc, there is a strong 

\subsection{Simulating Other Transients}
For the several other transients that could be in the field of view, (see Fig.~\ref{fig:mock_skymap}
for an example) we generate a representative sample of other possible objects. For this study we only
use extra-galactic transients: SALT2~\citep{Guy_2007} type Ia supernovae (SNe~Ia), core-collapse supernovae from the Non-negative Matrix Factorization (SNII-NMF) model,~\citep{plasticc} more recent core-collapse supernovae by~\cite{Vincenzi_2019}, (Vincenzi SNII), type Ibc supernovae, (SNe~Ibc), type Iax (SNe~Iax) supernovae, superluminous supernovae (SLSNe), and tidal disruption events (TDE). All these models are part of SNANA. These models, apart from the BU19 and the Vincenzi SNII model, were used in the PLAsTiCC challenge (see Sec.~4 and Fig.~1 from~\cite{plasticc}). The difference, however, was the use of Rubin Observatory wide-fast-deep observing strategy in case of PLAsTiCC.

Here, we regenerate these with the same ZTF observing conditions library derived from ZTF DR3 and used for the KNe as described earlier. In Appendix~\ref{appendix:example_lightcurves} we show some example lightcurves from the training set. We do not make additional cuts on the dataset to balance
the different types of objects. This choice is different from that used in MU19, as that worked focused on accurate multi-class classification, whereas our focus is on separating KNe from all other classes, which we regard as contaminants for the purpose of this work. While we use the BU19 and KN17 models and observing schedule of ZTF DR3 for demonstration here, the prescription can be easily applied to other lightcurve models, and future survey schedules, in particular Rubin Observatory and \emph{NGRST}. The choice made here is motivated by the potential coincident survey operations with the upcoming fourth LIGO-Virgo-Kagra observing run.

\begin{table}
	\centering
    \caption{The table lists the
    number count of different types of objects used for training}
    \begin{tabular}{cc}
    \hline
    Object Type & Number count \\
    \hline
    Bulla KN         & 5126 \\
    Kasen KN         & 3707 \\
    SALT2 SNIa       & 24670 \\
    SLSN             & 16926 \\
    SNII NMF         & 24232 \\
    SNIa 91bg        & 28620 \\
    SNIax            & 21686 \\
    SNIbc            & 21258 \\
    TDE              & 19359 \\
    Vincenzi SNII    & 19909 \\
    \hline
    \end{tabular}\label{tab:object_count}
\end{table}
The number count of the different types of objects are given in Table~\ref{tab:object_count}. It should be noted that this is a simplification, and more accurately, the number counts depends on the relative rates of explosive transients, and their detection efficiency. While the relative rates of contaminating extragalactic classes may affect classification, we can expect some contaminants to be more difficult to distinguish than others with only initial photometry. Here, we make a simple conservative approximation, and keep an order of magnitude more of the other objects compared to KN. This is motivated by our expectation that, while we are simulating events using ZTF's untargeted survey strategy, most efforts using the tool developed in this work will be applying it to images acquired through a dedicated targeted search, precisely because they are trying to identify the EM counterpart immediately after receiving a GW alert. While the KNe volumetric rate is roughly one-thousandth the local SN~Ia rate, given the context of a target-of-opportunity search for KNe, the number of KNe discovered will inevitably be disproportional to their volumetric rate. We have verified that the extreme case of repeating training with a roughly equal number of objects instead -- modeling a target-of-opportunity search that is unreasonably effective -- gives a similar result. In Sec.~\ref{sec:training} (see Table~\ref{tab:auc_vals}), we further repeat the training/validation process for different KN/non-KN object proportion and do not observe significant difference in performance.
The other extreme -- a machine learning algorithm conditioned on a training set that includes objects in proportion to their volumetric rate and detection efficiency -- is unsuitable for our goal of detecting kilonovae, but may be of use to groups looking to retrospectively classify all objects from a survey. It such a case if an imbalance is necessary for technical/practical reasons, the output needs to be re-weighted based on rates and detection efficiency of different object types.
% may lead to biased results in certain cases, for example, archival analyses aimed at multi-object
% classification for population inference. In such cases, if the training set is imbalanced, sometimes
% for technical reasons, 

\subsection{Simulating Sky-localizations}
\begin{figure}
    \centering
    \includegraphics[width=1.0\columnwidth]{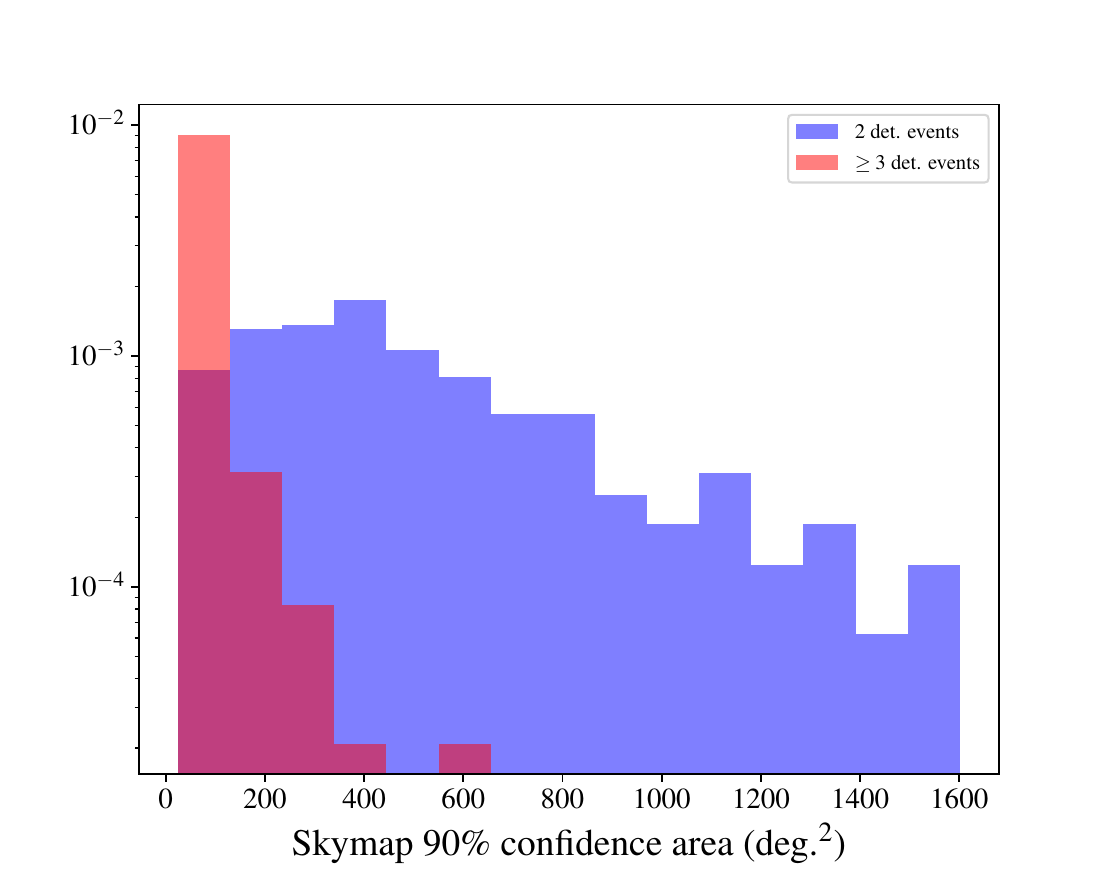}
    \caption{Distribution of the 90\% confidence regions of the 2-detector vs $\geq3$-detector events
             in the training set.}
    \label{fig:two_vs_three_det_skyreas}
\end{figure}
The GW alerts contain information about the merger time, how many ground-based interferometers
detected the signal, and the significance of the candidate. They also contain additional data products
that aid the hunt for any potential EM 
counterpart.~\footnote{\url{https://emfollow.docs.ligo.org/userguide/content.html}} 
% In the third
% observing run (O3), these included a 3D sky-localization probability
% map~\citep{singer_and_price, going_the_distance}, source properties of the binary -- whether
% the binary is expected to have remnant matter post merger~\citep{2020ApJ...896...54C}, and the
% low-latency astrophysical classification of the trigger~\citep{2020CQGra..37d5007K}. 
% Using such
% contextual information as features is important to exploit full information that is available
% at the time of discovery. 
For follow-up purposes, the sky-localization is of particular importance
since it determines scheduling and observing strategy. Additionally, the possibility of encountering
contaminants increases drastically with increasing sky-localization area. We compute and associate a GW skymap with each KNe lightcurve instance described in the previous section.

The basic principle of sky-localization utilizes the difference in the time of arrival of the signal
between the GW detector-network to localize the source. The GW signal depends on the intrinsic properties of the binary, like the masses, and spins of the binaries, and also extrinsic parameters like the distance and inclination with respect to the line-of-sight. The detectability of the signal, and the measurement uncertainties, like the localization area in sky, depends on the signal-to-noise ratio (S/N) produced at the detector. On the other hand, the amount of ejecta released from the aftermath of NS merger depends on the complex high-energy astrophysics of the merger environment, and the NS equation of state (EoS)~\citep[for example, see][for a review]{Shibata_2019}.
While the accurate answer to ejecta properties is given by numerical relativity
(NR) simulations, empirical fits to NR results have been performed to estimate the ejecta mass
given the binary source parameters and the NS EoS. Here, we use such a relation due
to~\cite{2017CQGra..34j5014D} (hereafter DU17) as implemented in the \texttt{gwemlightcurves}
library~\footnote{\url{https://gwemlightcurves.github.io/}} used in~\cite{coughlin_2020}.
The DU17 relation provides the ejecta mass given the binary parameters as,
\begin{equation}
    \mej = \mej(m_{1,2}, C_{1,2}),
    \label{eq:ejecta_mass_fit}
\end{equation}
where $m_{1,2}$ are the component masses and $C_{1,2} = (G/c^2)(m_{1,2}/R_{1,2})$ is the
compactness of each component. The compactness is a function of the equation of state (EoS)
of NSs. While there is still uncertainty in the true NS EoS, the tidal deformability
measurement from GW170817 data,~\citep{2018PhRvL.121p1101A} and the constraints on mass and
radius from NICER~\citep{2019ApJ...887L..24M} have ruled out several EoSs in literature.
Recently, \cite{2021arXiv210408681G} have performed model selection of the EoSs in the
literature against GW170817 data. For this study, we make a choice of the APR4\_EPP EoS, which
is well motivated physically, and well supported by GW170817 data.
%(see \cite{2021arXiv210408681G} Table~I \& II).
This is justified because the ejecta mass is not a strong function of the compactness
if we restrict to the set of EoS that have support from mass-radius measurements
from GW170817 or NICER. We show this in Appendix~\ref{appendix:mej_eos_variation}.
For each KN lightcurve, parameterized by $\mej$, we solve the DU17 relation to obtain the
component masses of the binary. We ignore spins for this analysis. Note that the solution
from $\mej \rightarrow m_{1,2}$ is not one-to-one. In
practice, we obtain the solution using a stochastic technique -- we sample uniformly in
chirp mass and mass ratio, obtain the component masses, and use Eq.~(\ref{eq:ejecta_mass_fit})
to match the ejecta mass value to the nearest BU19 SED $\mej$ grid-point.
We also use the $\thetaobs$ parameter as the inclination of the binary orbit with respect
to the line of sight. For the KN17 models, we follow the identical procedure to obtain
binary parameters from the ejecta mass, but associate a random inclination angle based
on a uniform prior on cosine of the inclination. In this manner we obtain the binary
parameters. 

We then use the binary parameters to generate GW strain using an analytic waveform
approximant,~\citep{PhysRevD.71.084008}~\footnote{TaylorF2 correct up to 3.5 post-newtonian order in
amplitude and phase} and compute the signal-to-noise (S/N) time series
in the HLVK detectors based on different noise realizations. We consider the projected sensitivity in the
LIGO/Virgo/KAGRA fourth observing run (O4) and the design-sensitivity expected in the fifth
observing run (O5). To generate the S/N time series, we use the \texttt{bayestar-realize-coincs}
tool from the \texttt{ligo.skymap} library.\footnote{\url{https://lscsoft.docs.ligo.org/ligo.skymap/}}
This simulates a match-filter search using gaussian-noise. We impose a threshold of S/N $\geq 4$ for a single
GW detector, a network S/N $\geq 12$, and a coincidence in at least two detectors to claim discovery in GWs.
Subsequently, we run the low-latency sky-localization code BAYESTAR~\citep{singer_and_price} on the
detected sources to obtain a skymap associated with each KN lightcurve. In Fig.~\ref{fig:two_vs_three_det_skyreas}
we show a comparison between the 90\% skyareas of the two-detector and greater than two-detector events.

\section{Training Data}\label{sec:training}
\subsection{Training set}
\begin{figure*}[p]
    \centering
    \includegraphics[width=0.45\textwidth]{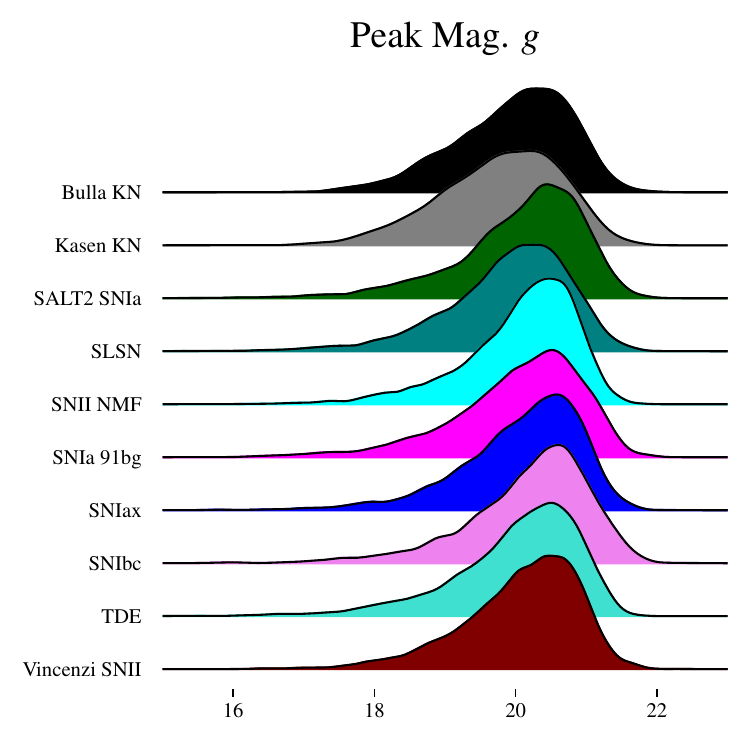}
    \includegraphics[width=0.45\textwidth]{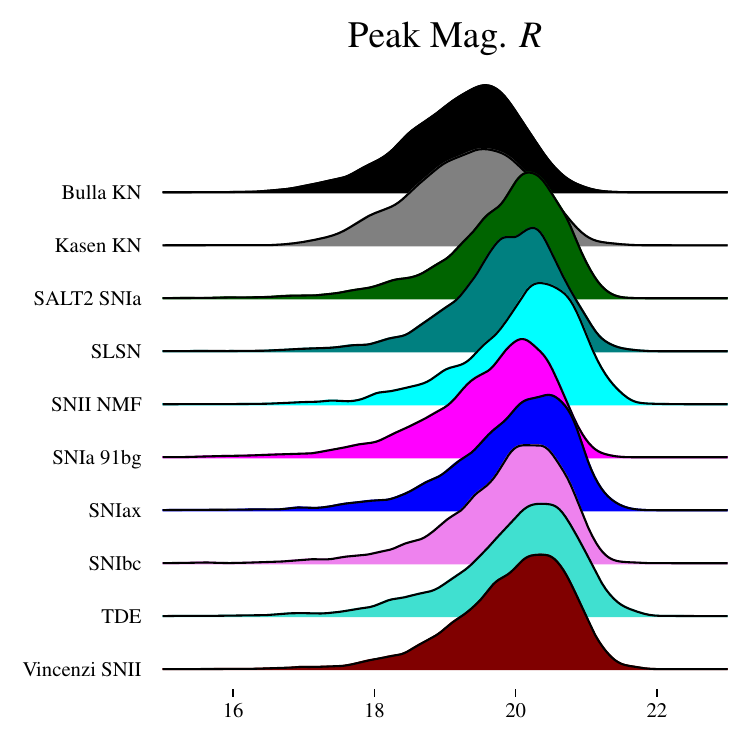}
    \\
    \includegraphics[width=0.45\textwidth]{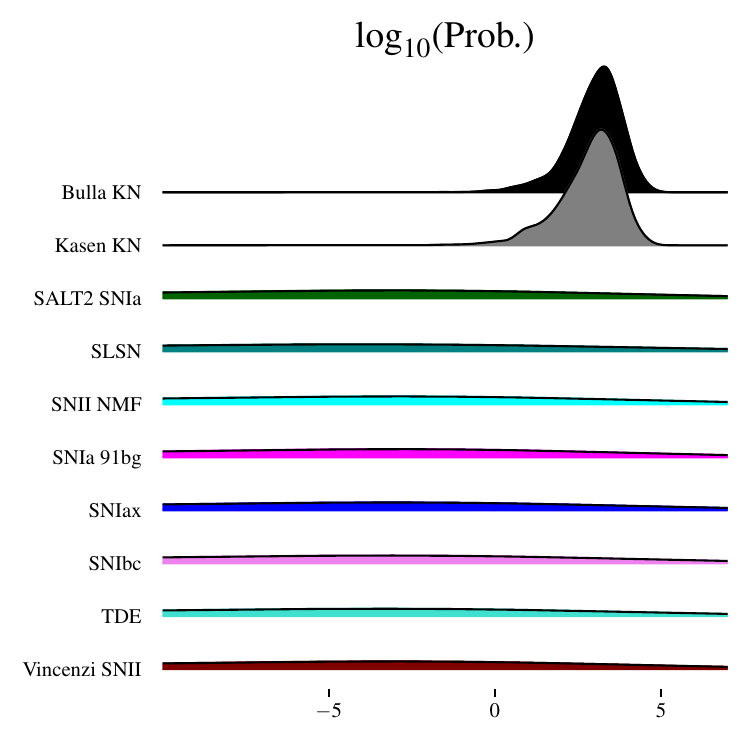}
    \includegraphics[width=0.45\textwidth]{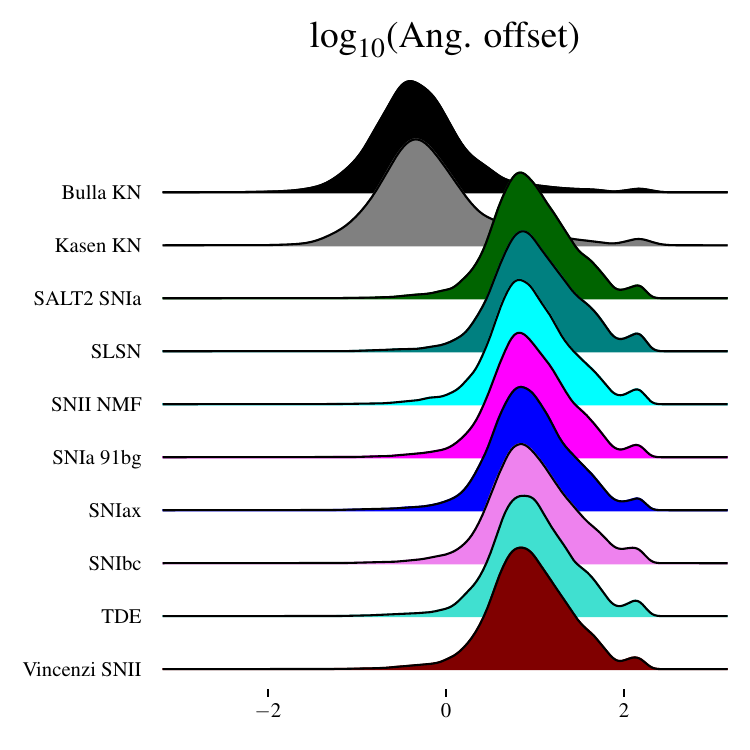}
    \caption{\textbf{Upper panel}: Peak $g$ and $R$ magnitude distribution. As
    expected, there is no preference for a particular object type in terms of
    the observed flux. \textbf{Lower panel}: Distribution of line-of-sight
    probabilities and angular offsets in degrees from the mode of the skymap
    }
    \label{fig:peakmag_logprob_distribution}
\end{figure*}
\begin{table}
    \centering
    \begin{tabular}{c|c}
    \hline
    \hline
    Fraction of non-KN objects & AUC at 3rd day after trigger \\
    \hline
    Complete set &        0.99970 \\
    0.50x non-KN objects & 0.99957 \\
    0.33x non-KN objects & 0.99950 \\
    0.25x non-KN objects & 0.99881 \\
    \hline
    \end{tabular}
    \caption{Area under the (receiver operator statistic) curve (AUC) at 3rd day after trigger.
    We progressively reduce the fraction of non-KN objects in the training set using a fraction
    of the total non-KN objects in the total set each time to test the impact of relative imbalance
    of objects towards binary classification. We observe that the relative imbalance does not
    affect the binary classification performance significantly.} \label{tab:auc_vals}
\end{table}
With each simulated kilonova light curve associated with a GW trigger alert and skymap, as well
as the simulated light curves of contaminants from multiple plausible astrophysical classes, we
preprocess the events and label events to assemble the training set as follows:
\begin{itemize}
    \item We assume the explosion time of the kilonova to coincide with the GW trigger. This is justified
    since the time to black-hole formation from the merger of neutron stars is $\mathcal{O}(10-100)$
    (milliseconds; see~\cite{Radice_2018} for example), at which point the dynamical ejecta is expelled.
    \item For each kilonova in our sample, we consider any contaminant light curve in our simulated sample
    that has a first detection within 7 days of the GW trigger. With the dense set of simulations, this step
    yields $\mathcal{O}(100-1000)$ objects, most of which have high angular separation from the high probability
    density region in the GW skymap corresponding to the kilonova.
    \item We rank these objects based on their line-of-sight probability, and select the top 20
    objects for each KN/skymap combination. Thus, our training set contains about 20x the
    number of KN lightcurves we start out with. For this study we use about $8\times 10^3$
    KN/skymaps combinations, which implies that our training set has $\sim 1.6\times10^5$
    total lightcurves/contextual information combinations. We have verified that
    the performance is not affected by changing the proportion of KN vs. non-KN objects in the
    training set. In Table~\ref{tab:auc_vals}, we show the area under the receiver operator statistic
    curve on the 3rd day after trigger by repeating the training/validation changing the proportion
    of KN vs non-KN objects each time. We do not observe a significant difference in performance.
    \item The KN (BU19 or KN17) are labeled as ``Kilonova'' while the other objects
    are labeled as ``Other'' class.
\end{itemize}

In the upper panel of Fig.~\ref{fig:peakmag_logprob_distribution}, we show
the distributions of the $g$ and $R$ observed band fluxes as output by SNANA. As expected, since all the events are simulated using the same survey properties, there is no significant bias that can discriminate any specific object type. While the distribution of peak absolute magnitudes and colors can differ between the classes, the observed magnitude distributions are largely similar as they are determined by the survey capabilities. With only sparse early-time photometric data, it would not be possible to discriminate between kilonovae and other classes effectively. However, in the bottom panel we show the distribution of the line-of-sight probability and the angular offset from the mode of the GW skymap for all the objects. These distributions sets the KNs apart from the other classes, as the GW skymap is associated with the kilonovae, whereas any association between the skymap and a light curve from some other class of object will arise purely by coincidence. When considering a population of contaminating events, no correlation will persist between the GW skymaps and the light curves. This highlights the value of considering contextual information in addition to the light curve, particularly for early classification, even if the contextual information is drawn from a different survey.

\section{Binary Classification}\label{sec:neural-net}
\begin{figure}
    \centering
    \includegraphics[width=0.8\columnwidth]{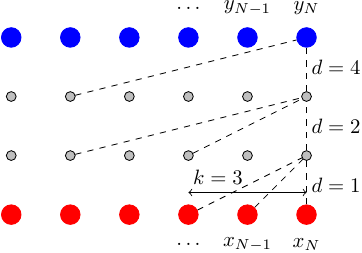}\\
    \includegraphics[width=1.0\columnwidth]{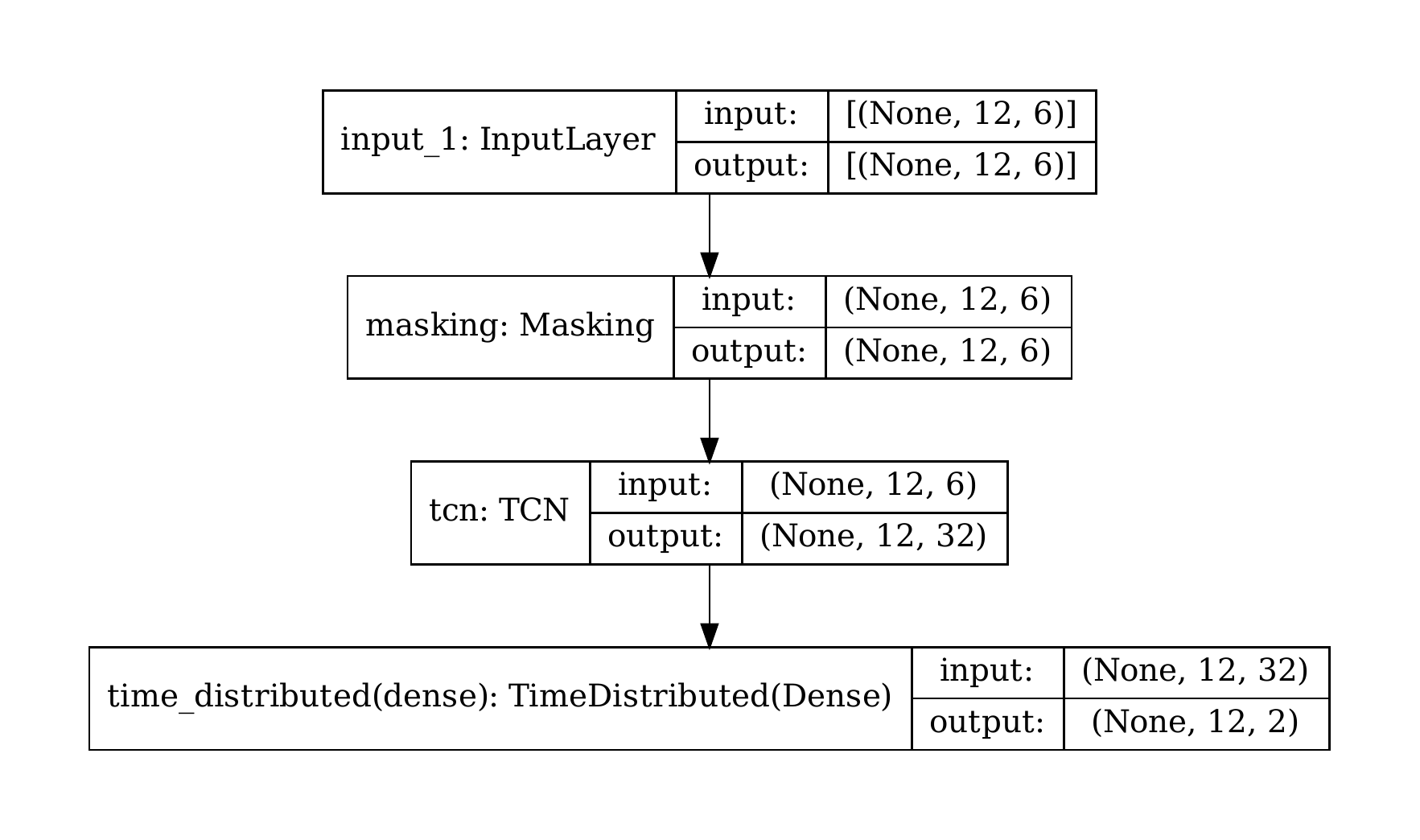}
    \caption{\textbf{Upper panel}: A basic architecture of a TCN, reproduced
    based on Fig~1. of \citet{tcn} (see for more details). The sketch shows
    the causal nature of the convolutions -- output $y_N$ depends on
    the input till $x_N$. The \emph{filter} length, $k$, and the
    dilations, $d$, control the preservation of history of the
    time-series. \textbf{Lower panel}: Layers implemented using \texttt{keras},
    showing input and output shape at each step.
    }
    \label{fig:basic_tcn}
\end{figure}
We use a similar neural network architecture as RAPID. While the original architecture
used gated recurrent units (GRUs), we use a temporal convolutional network (TCN)
architecture~\citep{tcn} as implemented in  
\texttt{keras}~\citep{KerasTCN}.\footnote{\url{https://github.com/philipperemy/keras-tcn}}
As mentioned earlier, here we are only concerned whether an object is/is not a KN. Thus, we
modify the architecture used in MU19 to a binary classifier to act as a filter for KNe.\footnote{
At the time of publication, this is a separate branch of the RAPID codebase.}

\subsection{Temporal Convolutional Network}
A TCN architecture \cite{tcn} uses two basic principles -- 1) the output is of the same length as the input, and
2) the internal convolutions are \emph{causal}. To illustrate this point, consider a 
sequence, $\{x_1, \dots, x_N\}$; the output is a sequence $\{y_1, \dots, y_N\}$, and the
convolutions of the hidden layers of the architecture are such that the output, $y_n$ only depends
on $\{x_1,\dots,x_n\}$, where $n \leq N$. We show a basic sketch in Fig.~\ref{fig:basic_tcn}.
While TCNs have previously been used in the literature to model periodic variable stars,
\citep[see][for example]{2020arXiv200507773M} this is the first use of this architecture to
classify astrophysical transients.

% what does this mean - temporal long-term history
There are two ways to preserve temporal long-term history of the
input sequence: using larger \emph{filter}~\footnote{called ``kernel'' in the
{\tt{keras}} nomenclature} length, and using \emph{dilated} layers of causal
convolutions in this architecture. Increasing both correspond to extraction of
features that are correlated farther in time. In our case, the temporal
data is the flux of the object as a function of time i.e., the lightcurve. Additionally, the contextual
information is also an input, but its value does not change with time, for example, the angular offset
of the object from the mode of the skymap, or its line-of-sight probability.
Following MU19, we sample the lightcurve uniformly in $N$ steps. Our input
matrix, therefore, has dimensions of $N\times(p + k)$, where $p$ is the number of passbands (3 for ZTF),
and $k$ is the number of contextual information features. In this case, this includes the
line-of-sight probability, the angular offset, and the 90\% area of the skymap.

The light curves are preprocessed before training. The flux for any
object is approximated using linear interpolation in between two observations, and
then sampled at regular intervals to get the input time-series, $\{x_1,\dots,x_N\}$. In order to
find the regular intervals, we consider the 80 percentile value of $\sim 0.7$ days as the time
difference between two successive observations of our simulated light curve.
While this is an empirical choice and can be changed, for the cadence of observations used by ZTF, using a much larger interval decimates the time variability, while using a shorter interval causes jitter due to interpolating in an interval with sparse data. We only consider the light curve flux within 10 days from the GW trigger for all simulated object. This limit arises from the physics of kilonovae itself. Even the most intrinsically bright kilonovae evolve over $\sim 1$ week after the BNS merger. If an EM counterpart cannot be identified within 10 days, we simply flag the event as failed, as kilonovae become too faint at optical wavelengths for any follow-up studies that begin after 10 days. This choice differs from that used in MU19, which focused on multi-class classification rather than kilonova counterpart identification. Within this 10-day window, we interpolate the flux linearly onto an evenly spaced $N = 12$ grid. This choice satisfies 
$0.7 \times 12 \gtrsim 1$ week, giving us sufficient time resolution to incorporate observations from most time-domain surveys with cadences of $\gtrsim1$~day, while also covering the KN evolution. Thus, our input matrix has a shape of $12 \times 6$. A summary of the dimensions of the input, output, and hidden layers in  the bottom panel of
Fig.~\ref{fig:basic_tcn}.

\subsection{Training}
\begin{figure}
    \centering
    \includegraphics[width=0.8\columnwidth, trim=1.1cm 0cm 0cm 0cm]
        {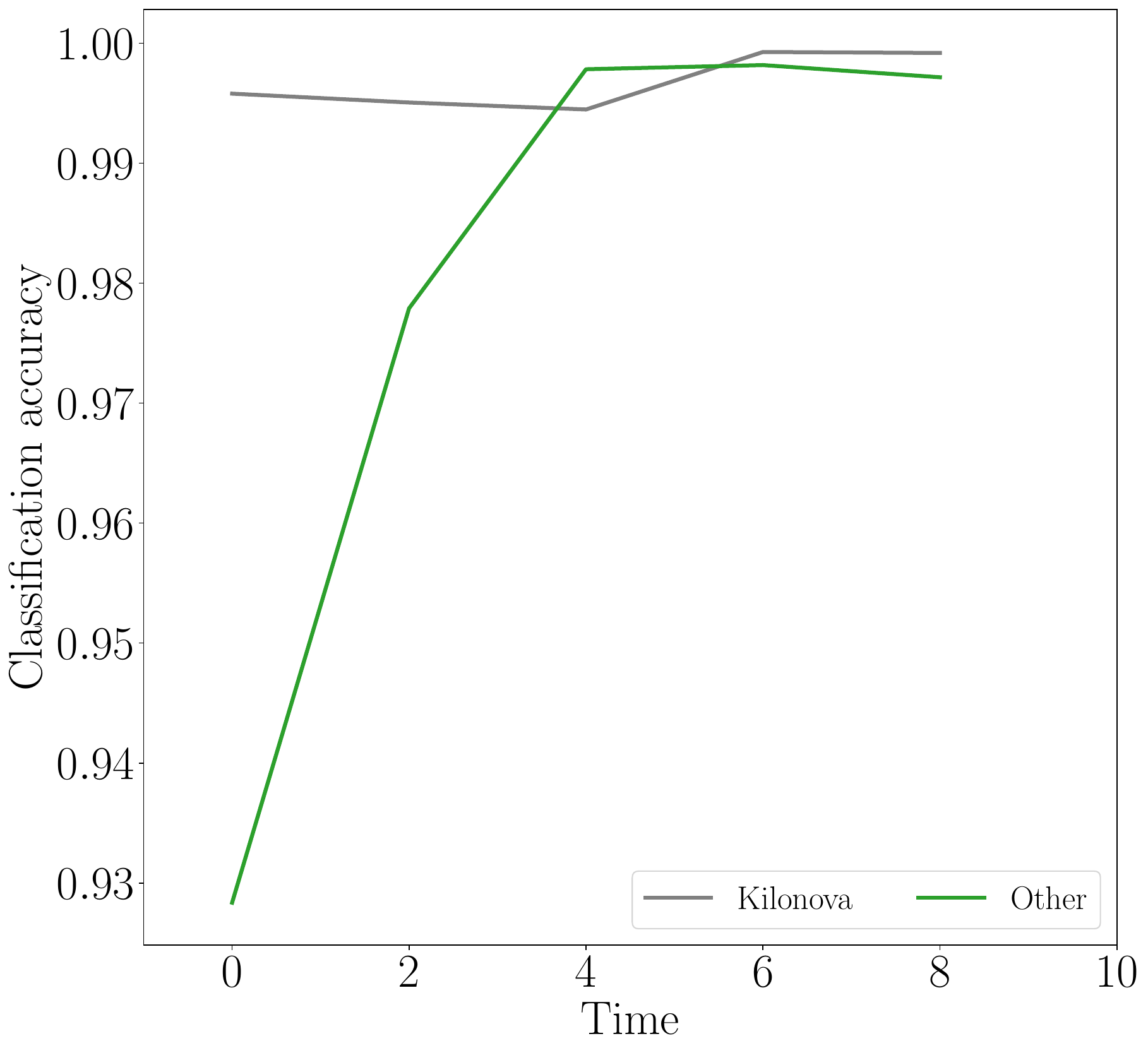}\\
    \includegraphics[width=0.94\columnwidth]
        {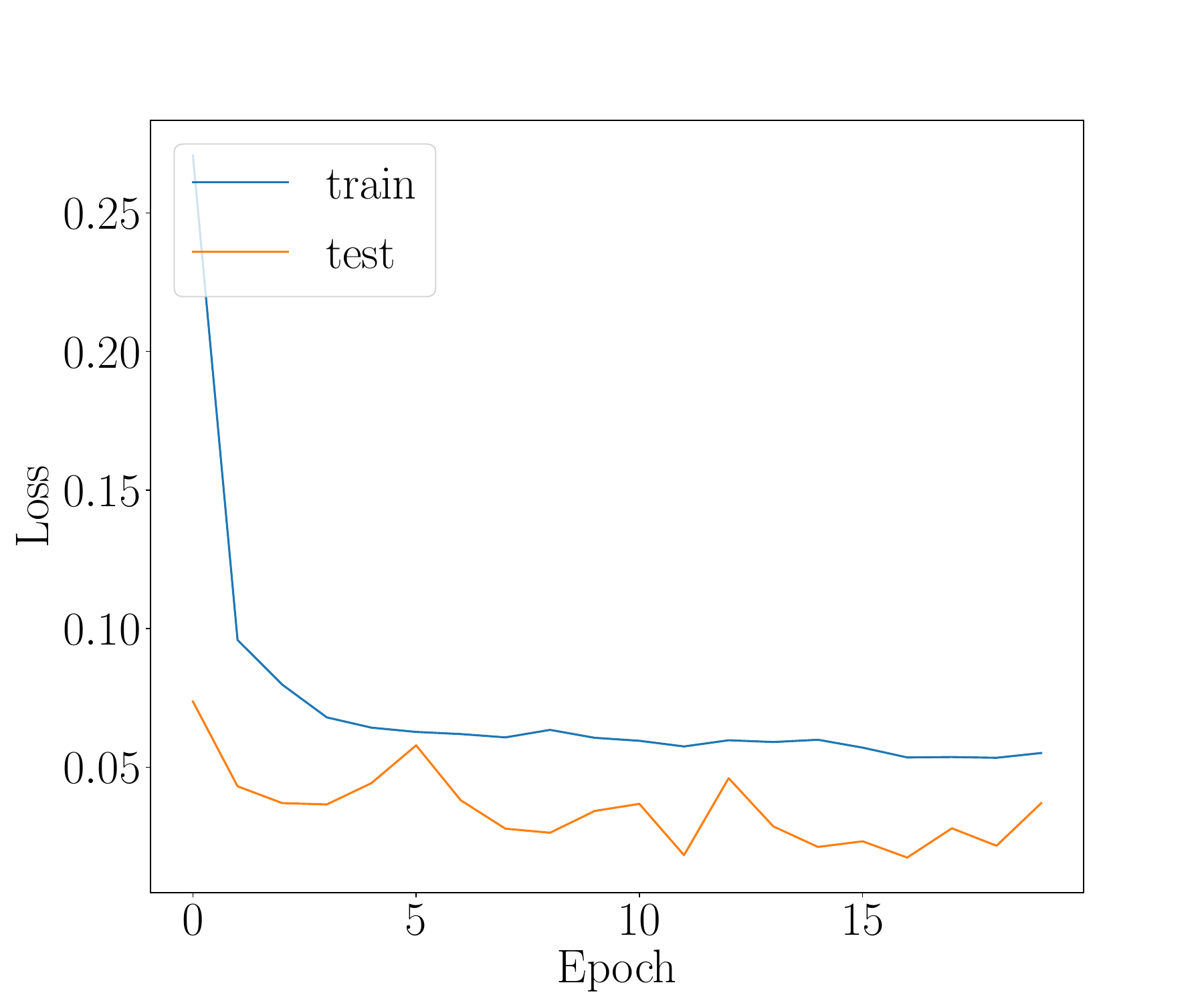}
    \caption{\textbf{Upper panel}: The accuracy of prediction with time for the
    binary classifier. \textbf{Lower panel}: The loss as a function of the training
    epoch.}
    \label{fig:model_loss}
\end{figure}
The kernel size of the hidden layers control the number of trainable parameters in
the network. In addition, from going from one hidden layer to the next, the input from
the previous layer is passed through a \emph{residual block}~\citep{2015arXiv151203385H}. This is
done to avoid vanishing gradients during training. In this step the input from the previous hidden layer
is passed through two causal convolutional layers and a dropout before being added
to itself. This further increases the number of trainable parameters. For our network, we use
a filter length, $k=2$, and dilation layers, $d = \{1, 2\}$. We use a kernel size of 32 for the
hidden layers (not shown in Fig.~\ref{fig:basic_tcn}). We also use 2
stacks~\footnote{\texttt{nb\_stacks=2} in \texttt{keras-tcn}.} to increase the
depth of the network. The total number of trainable parameters for the network
is $\sim 13000$. We use a dropout rate of $5\%$ to prevent overfitting. We use the categorical cross-entropy
as the loss function, and the Adam optimizer. We find that $\sim 20$ epochs of training
($\sim 5$ minutes total on 8 Intel core-i7 processors) is sufficient to minimize the loss. We use
$60/40\%$ of the data for training/validation fraction for our dataset. In Fig.~\ref{fig:model_loss}, we
show the accuracy per class with time, and loss as a function of the training epoch. 
It should be observed
that for most of the KNe, their sparse photometric data implies that the class is estimated solely
from their contextual information i.e., consistency with the GW skymap. For the other objects, however,
the photometric information is responsible for classifying them correctly i.e., we see the accuracy
increase with time. In either case, the $\geq 99\%$ accuracy is reached after $\sim 3$ days
from trigger. 

We would like to mention that the end-to-end dataset preparation
i.e., simulating binaries, running BAYESTAR to create skymaps, and using SNANA to prepare realistic
lightcurves, is the computationally expensive step in this work. We find that the computational cost in
training the neural network is significantly less, not requiring hardware optimizations, like
the use of GPUs, at this stage. We attribute this to the short duration photometry
i.e., the input arrays are relatively small compared to that needed for supernova lightcurve
classification as in MU19. Also, we only use certain features of the skymap to capture
the correlations between skymap and the location of other objects, instead of using the complete
HEALPix map as a feature vector. However, certain hardware optimizations, like GPU support,
may be done for future scalability since the underlying infrastructure of \texttt{tensorflow}
natively supports GPUs.
In the next section, we further validate the performance on simulated data that was not included
in training, as well as the real observations of GW170817, through similar passbands, but at different
facilities than our simulated ZTF training set.

\section{Results on Unseen data}\label{sec:results}
% \begin{figure}[t]
%     \centering
%     \includegraphics[width=0.7\columnwidth]{kn-prediction-results.png}
%     \caption{Predictions on all KN lightcurves used for training}
%     \label{fig:all_predictions}
% \end{figure}
% \subsection{Predictions on all lightcurves}
% In this section we show the results of predictions on all KN lightcurves
% along with the spread. In Fig.~\ref{fig:all_predictions}, we show the results.
\begin{figure*}
    \centering
    \includegraphics[width=0.32\textwidth]{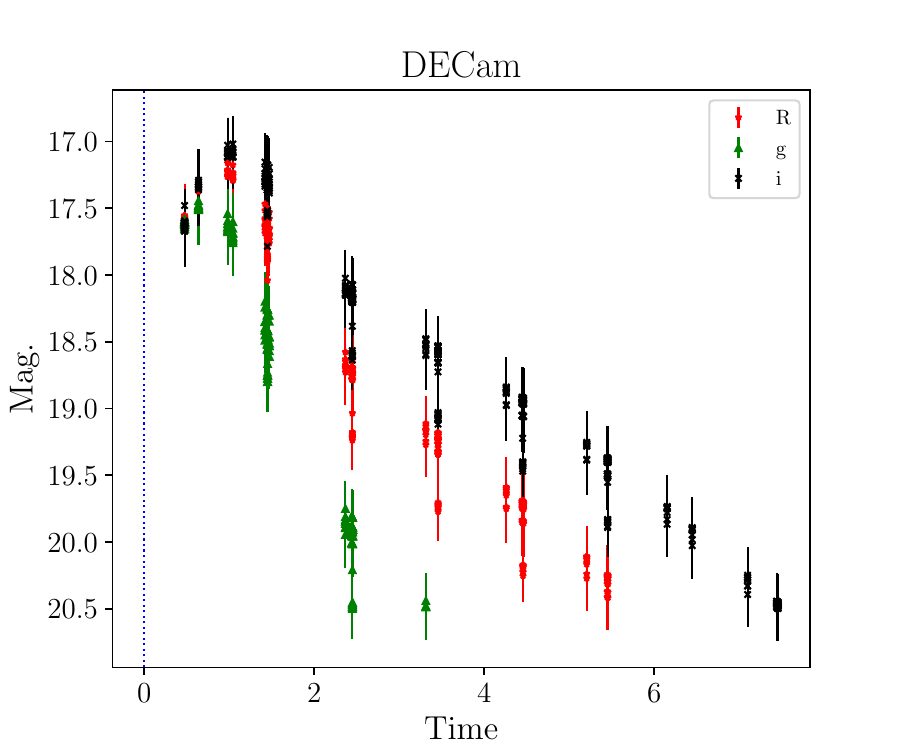}
    \includegraphics[width=0.32\textwidth]{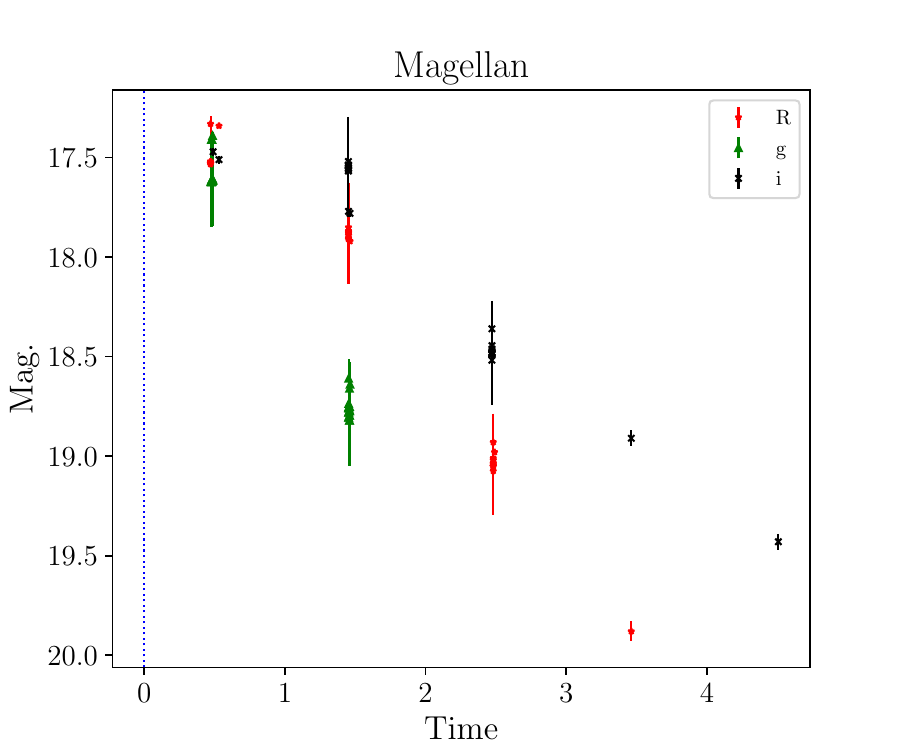}
    \includegraphics[width=0.32\textwidth]{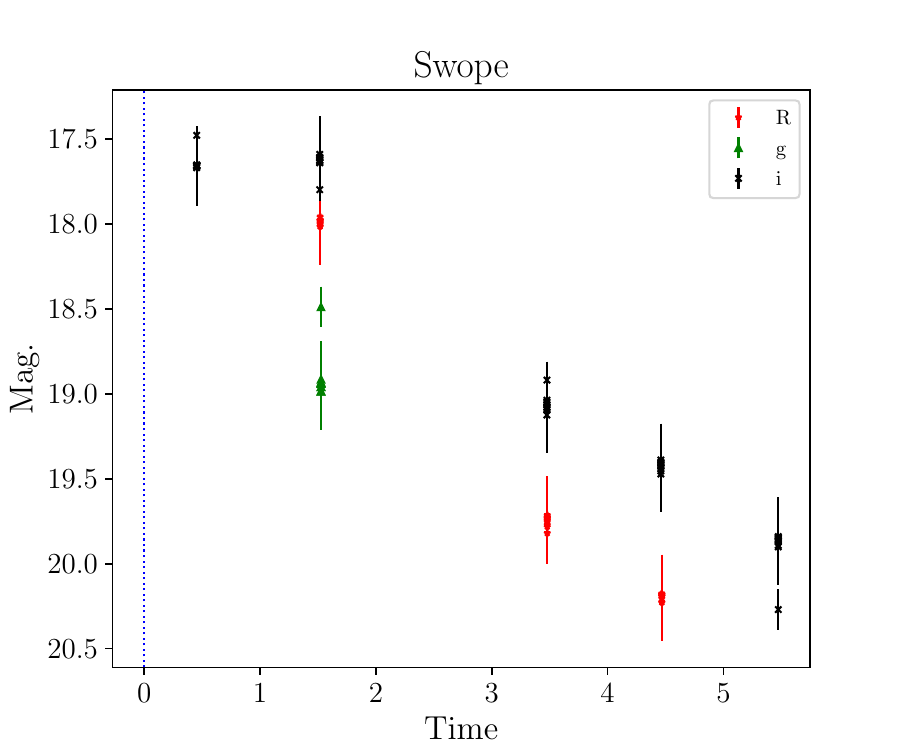}
    \\
    \includegraphics[width=0.32\textwidth]{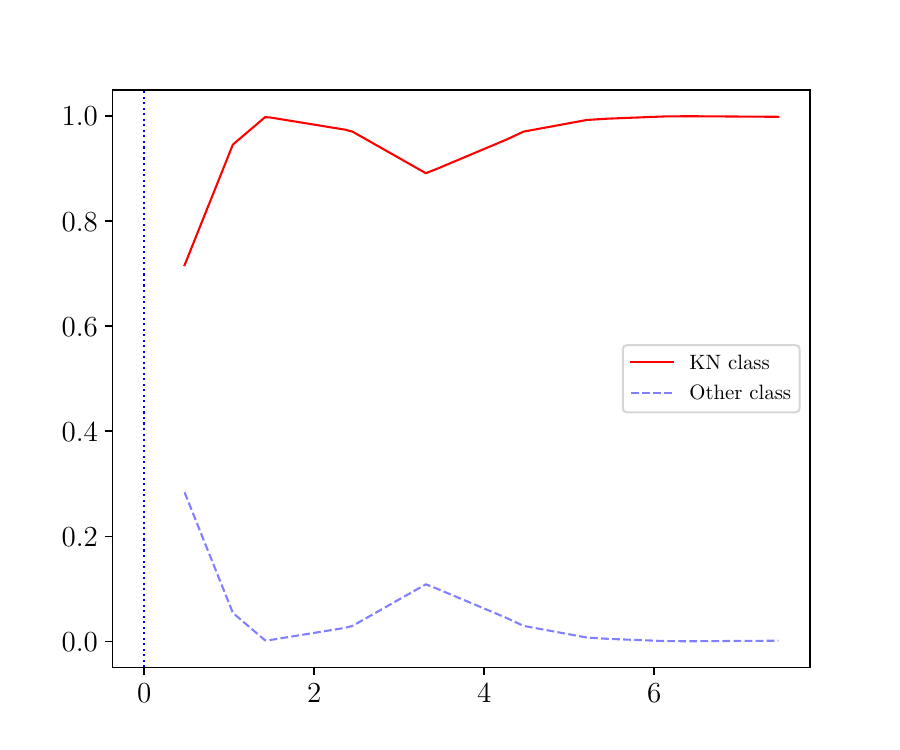}
    \includegraphics[width=0.32\textwidth]{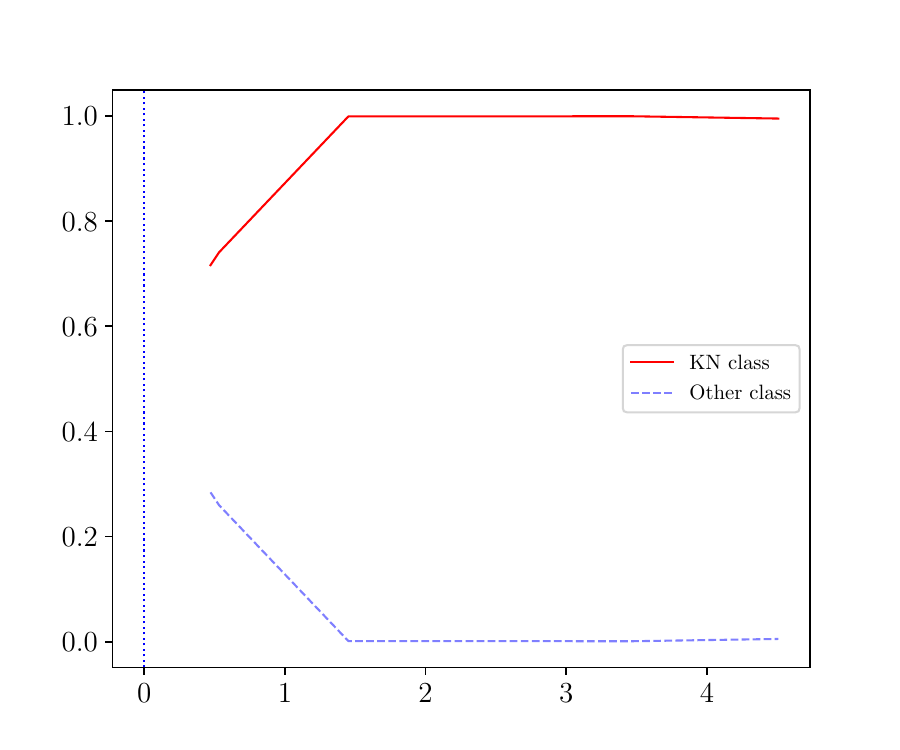}
    \includegraphics[width=0.32\textwidth]{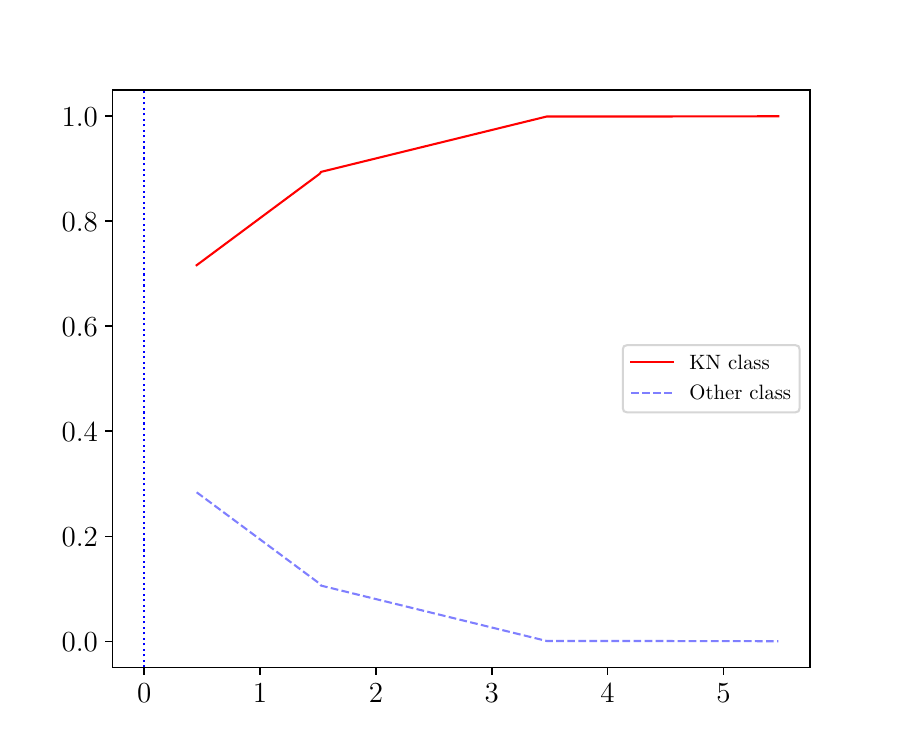}
    \\
    \includegraphics[width=0.32\textwidth]{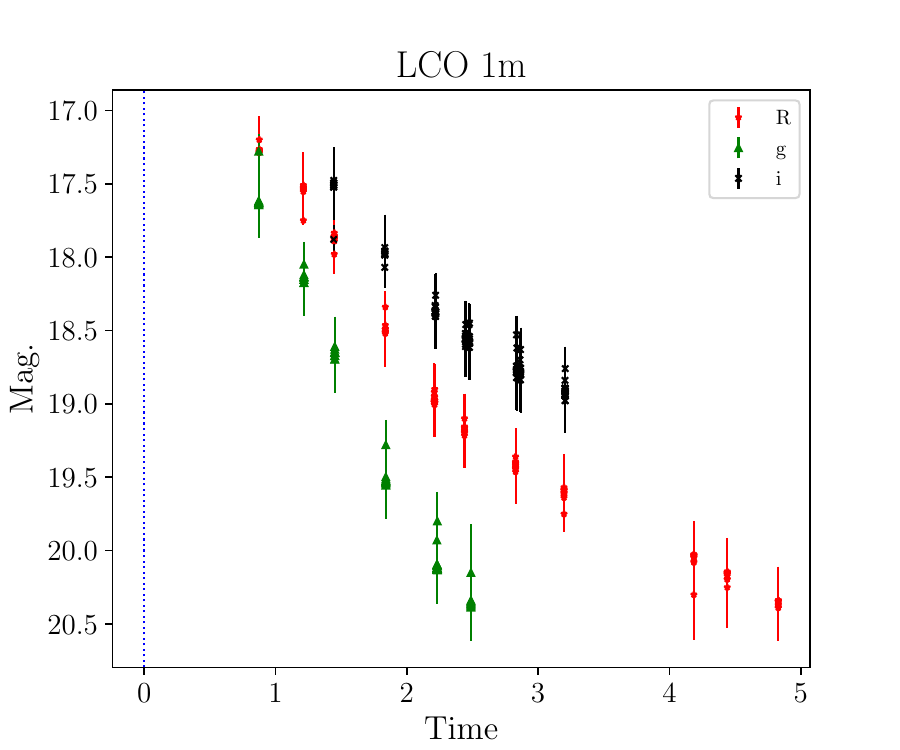}
    \includegraphics[width=0.32\textwidth]{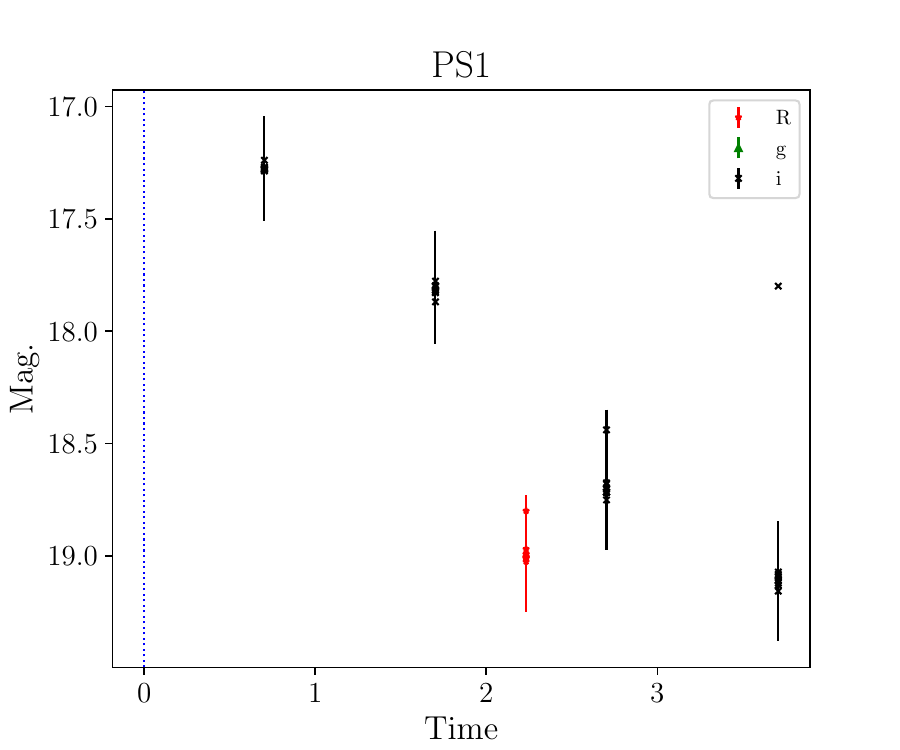}
    \includegraphics[width=0.32\textwidth]{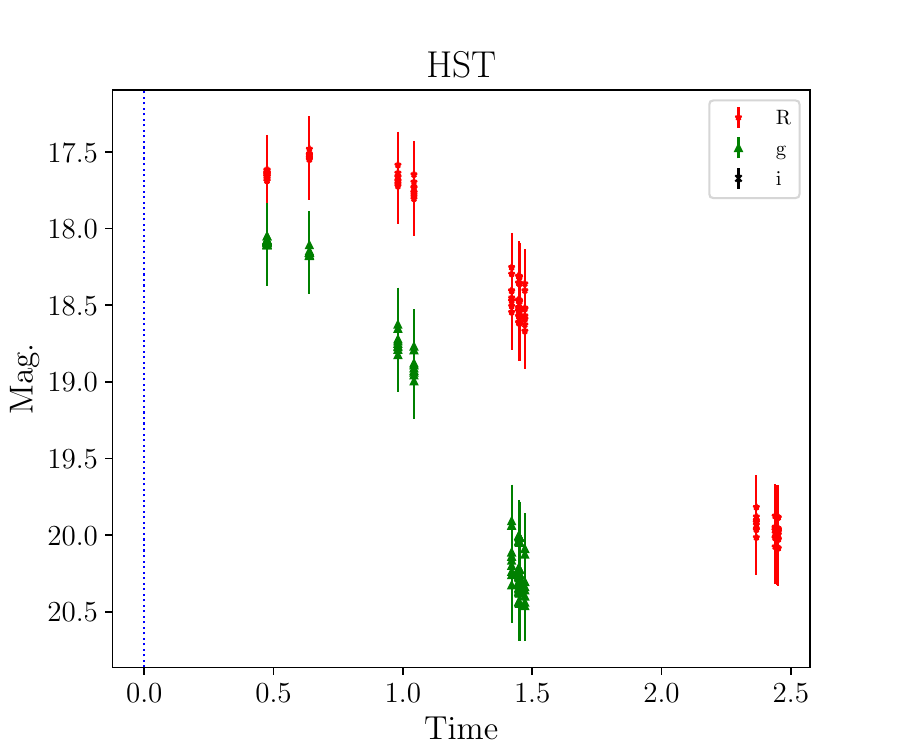}
    \\
    \includegraphics[width=0.32\textwidth]{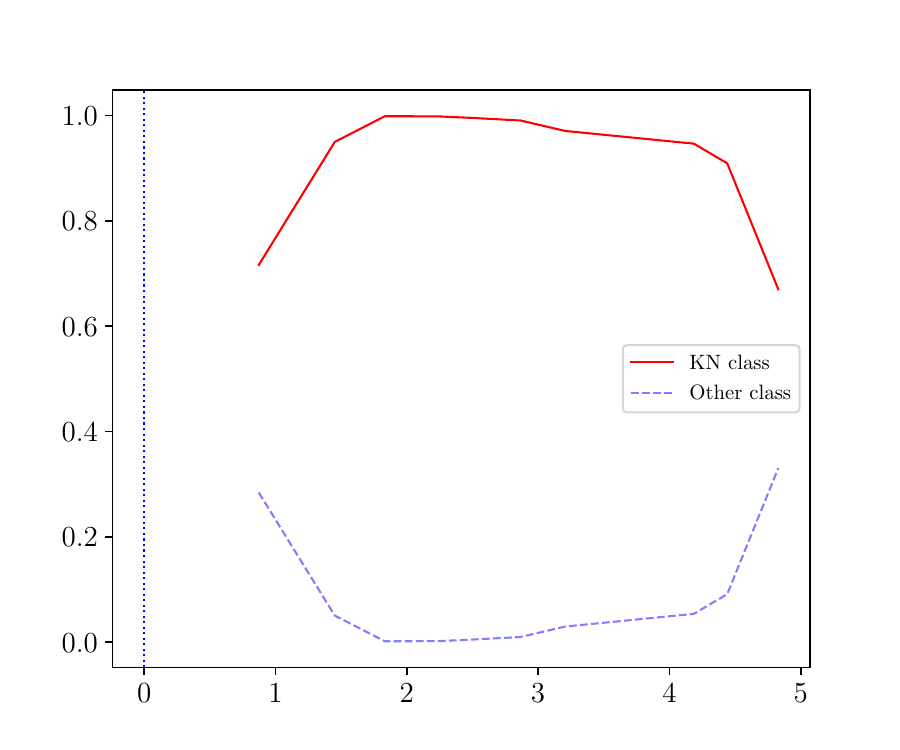}
    \includegraphics[width=0.32\textwidth]{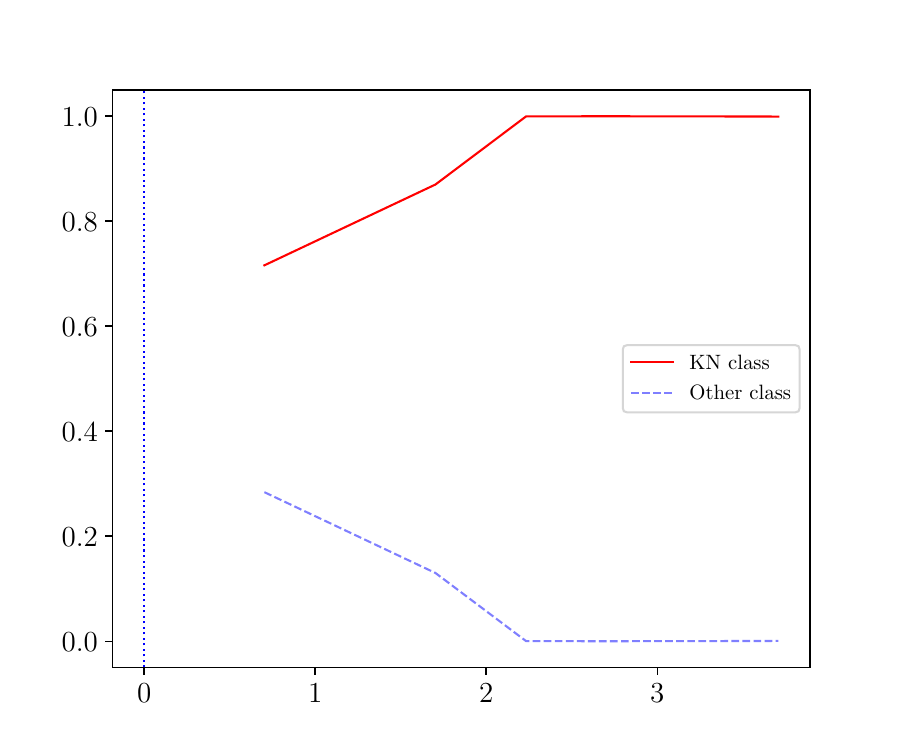}
    \includegraphics[width=0.32\textwidth]{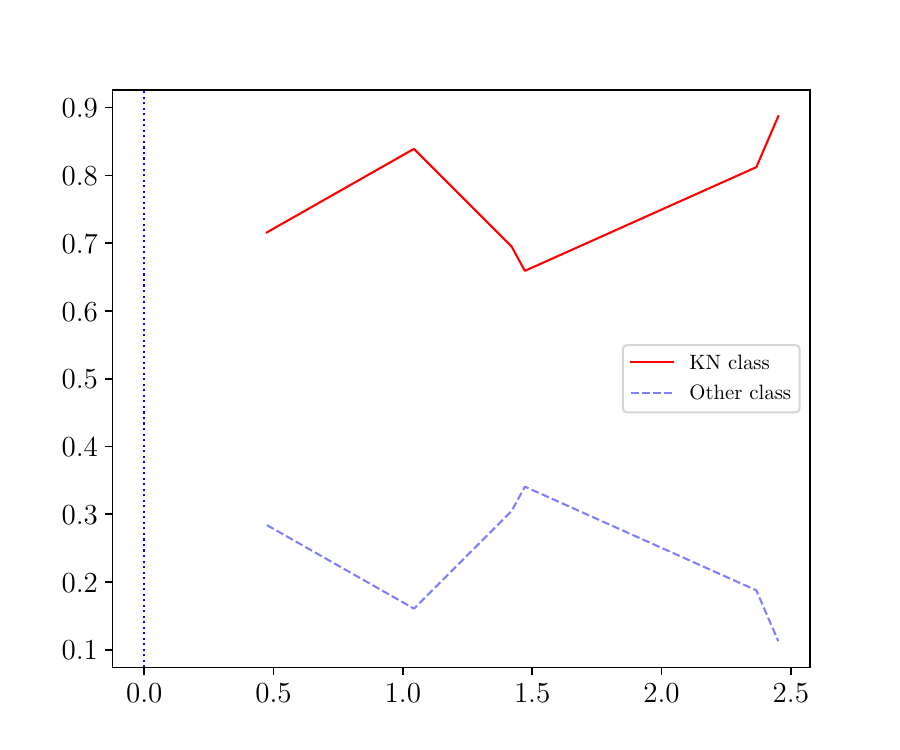}
    \caption{Predictions of the classifier on $\sim 1$ week data for GW170817
    obtained from the Open Kilonova catalog.}
    \label{fig:predictions_GW170817}
\end{figure*}
\subsection{Performance on GW170817}
The most natural choice to test the performance of El-CID is AT~2017gfo,
the kilonova from GW170817. We use the publicly
available data from the Open Kilonova Catalog.~\footnote{\url{https://kilonova.space/}} This dataset is a compilation
of the photometry from the various instruments that participated in the follow-up
of AT~2017gfo, and was not a part of our training sample. In Fig.~\ref{fig:predictions_GW170817},
we show the data for the first few days after the trigger for a subset of instruments
with optical passbands, similar to what we have used. We do not perform any
$k$-corrections to the magnitudes. We only consider the observation in filters used for training
-- $g$, $R$, $i$. Therefore, the data is extremely heterogeneous involving
different cadence and instrument properties.

Upon passing this to El-CID, we find that KN class is weighted strongly at the first detection, which is expected since AT~2017gfo lied in a high confidence region of the sky-localization.
We find that irrespective of the instrument, the classification stays correct. This shows that the classifier is robust to cadence and the temporal features are picked out instead. 

In some cases, like the LCO-1m case, the there is a decrease in KN class score at larger time intervals - this is a known feature. While most of the KNe in the training sample have observations in under a week from trigger, the LCO-1m lightcurves stay above threshold beyond a week as they were observed with exceptionally long exposures. These light curves therefore show a tendency of a drop in KN score in the later epochs as the real long-exposure observations are not similar to our training data which use a constant exposure as would be employed by an untargeted search. A possible solution to be considered in the future is the augmentation of the training set by including simulations with either a mix of small and large aperture telescopes, or by incorporating short and long duration exposures. 

In Appendix~\ref{appendix:design_sensitivity}, we also show the results from a classifier trained on $\geq 3$-detector coincidences in the design sensitivity era. We find that the results are not significantly affected. This is expected because the detection of the joint sources are mostly limited by the sensitivity of the telescope for future observing eras.

\subsection{Performance on AT~2019npv}
\begin{figure}
    \centering
    \includegraphics[width=0.95\columnwidth, trim=0cm 0.7cm 0cm 0.7cm, clip]
        {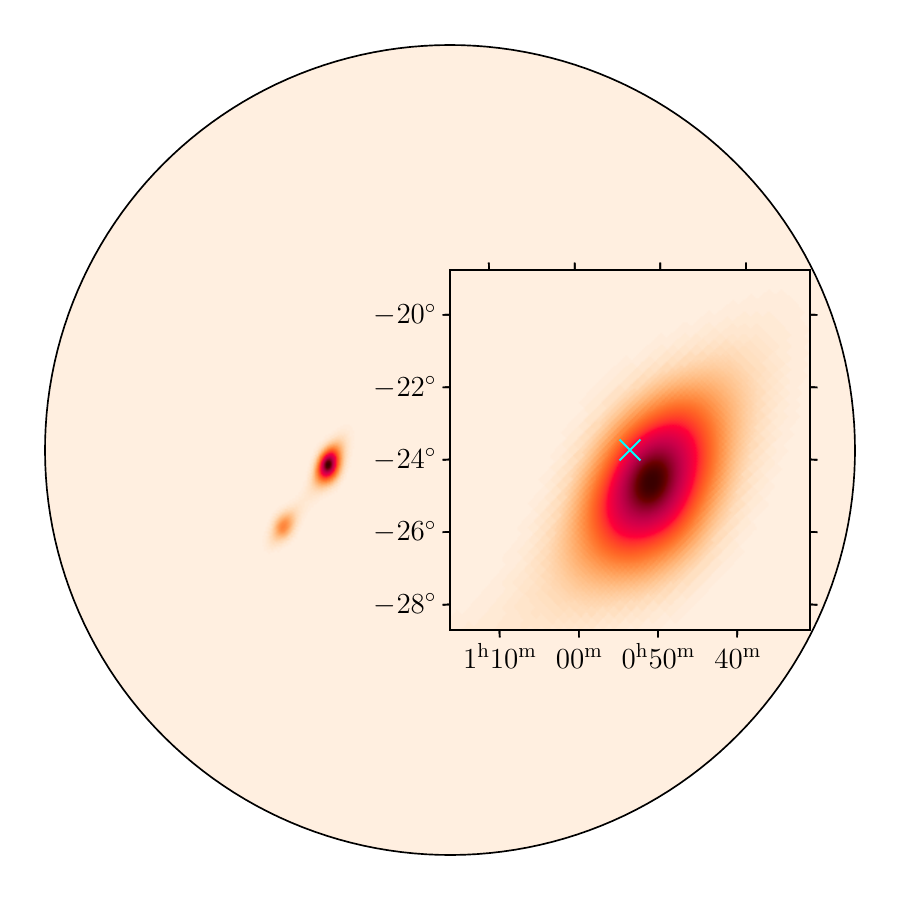}\\
    \includegraphics[width=1.0\columnwidth, trim=0cm 0cm 0cm 1cm, clip]
        {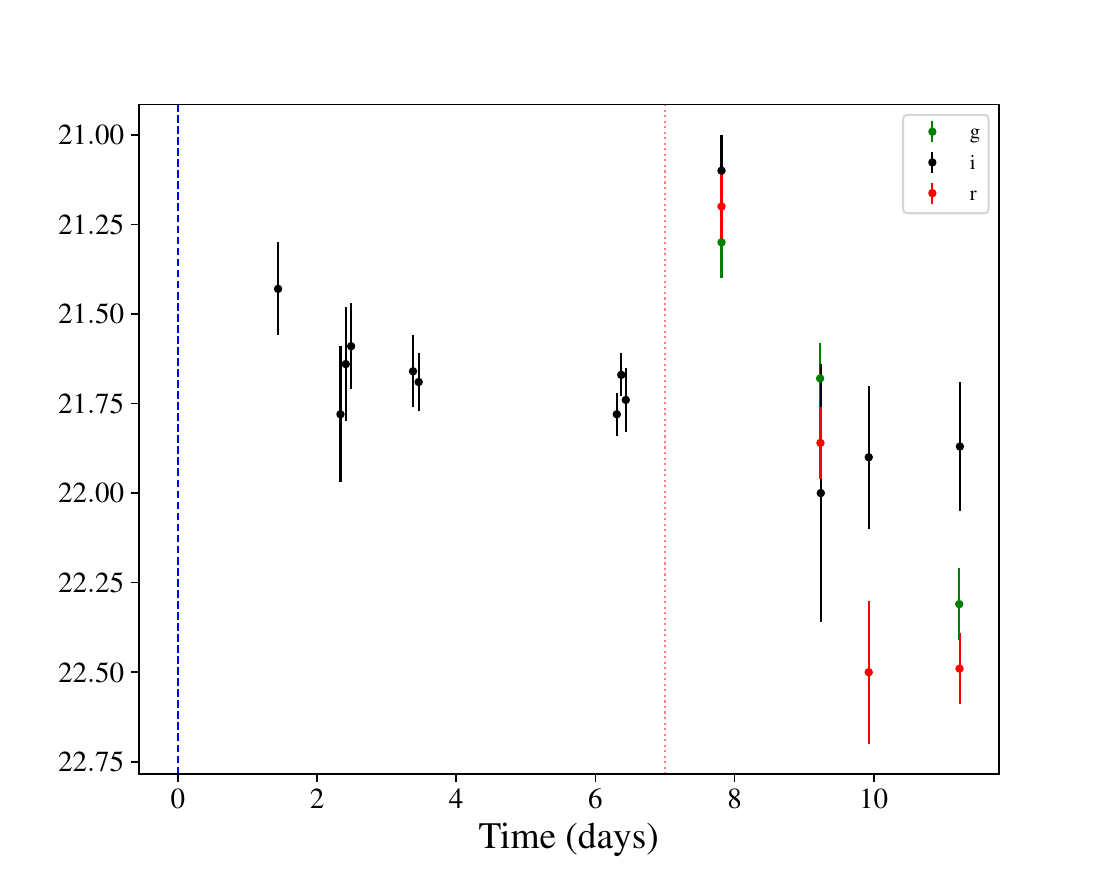}\\
    \includegraphics[width=1.0\columnwidth, trim=0cm 0cm 0cm 1cm, clip]
        {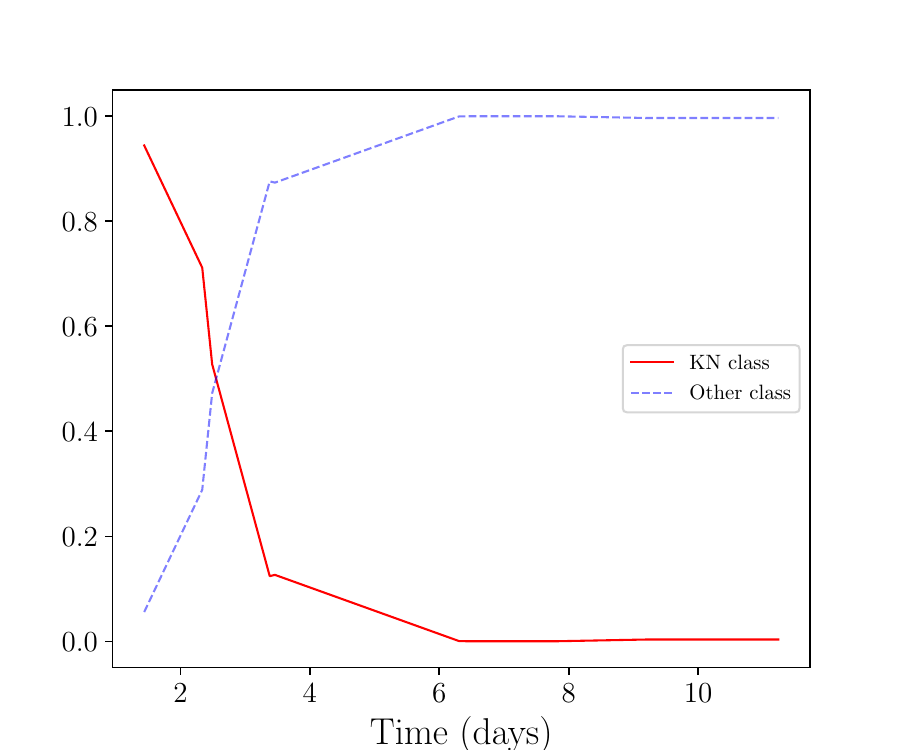}
    \caption{\textbf{Upper panel}: The skymap of the GW190814 event. The cross
    shows the position of AT~2019npv, a SN~Ibc. The object is in a high-confidence
    region of the skymap, and was considered a KN doppleganger few days into discovery
    \citep{Andreoni_2020}. \textbf{Middle panel}: The photometry of AT~2019npv taken
    from \citet{Andreoni_2020}. The reference time is the merger time of GW190814. The
    observation till $\sim 7$ days (dotted vertical line) is from DECam follow-up.
    Subsequent observations are from several observatories.
    \textbf{Lower panel}: El-CID results on AT~2019npv. We find, as expected, the
    initial contextual information weights the KN class strongly. However, after a
    few epochs of observation, the KN prediction drops.
    }
    \label{fig:at2019_npv}
\end{figure}
Next, we consider the performance on a type Ibc supernova, AT~2019npv, that was found
in a region of high-confidence in skymap of GW190814~\citep{Abbott_2020}. We show the
skymap and the location of the transient in Fig.~\ref{fig:at2019_npv}. GW190814 was the
best localized event during LVC run O3, with a preliminary skymap localization area of $\sim 20$
sq. degrees. This event was interesting for follow-up since it was reported to have
strong chance of having a NS component (mass less than 3 $M_{\odot}$~\citep{GCN25333}.
Several teams across the globe participated in the follow-up effort of this 
event~\citep{Antier_2019,Dobie_2019,Andreoni_2020,Morgan_2020,Vieira_2020,Watson_2020,de_Wet_2021}.
AT~2019npv was discovered by \cite{GCN25393} and confirmed by several groups
~\citep{GCN25398, GCN25455, GCN25457, GCN25474, GCN25485}. The redshift of this source was
consistent with the distance information from the skymap. We show the observations taken from~\cite{Andreoni_2020}
in the middle panel of Fig.~\ref{fig:at2019_npv}. The initial $i$-band observations in the first
week were taken with DECam. The later observations are from a mix of several 
instruments at different facilities. The data is taken from Fig.~3 of \cite{Andreoni_2020}. 

We find that as expected, the initial KN score is high due to the consistency between the event location with the high-confidence region of the GW skymap. However, every subsequent epoch of observation drives the KN score down -- as El-CID sees more of the evolution of the light curve, the contextual information carries less weight and the algorithm becomes more certain that this event is not a kilonova.

This serves as another verification of the classifier performance on unseen real data.
In Appendix~\ref{appendix:design_sensitivity}, we also show the results from using a
training set in the design sensitivity era.

\subsection{Performance on mock KN lightcurves in Rubin Observatory Era}
\begin{figure*}
    \centering
    \includegraphics[width=1.0\textwidth]{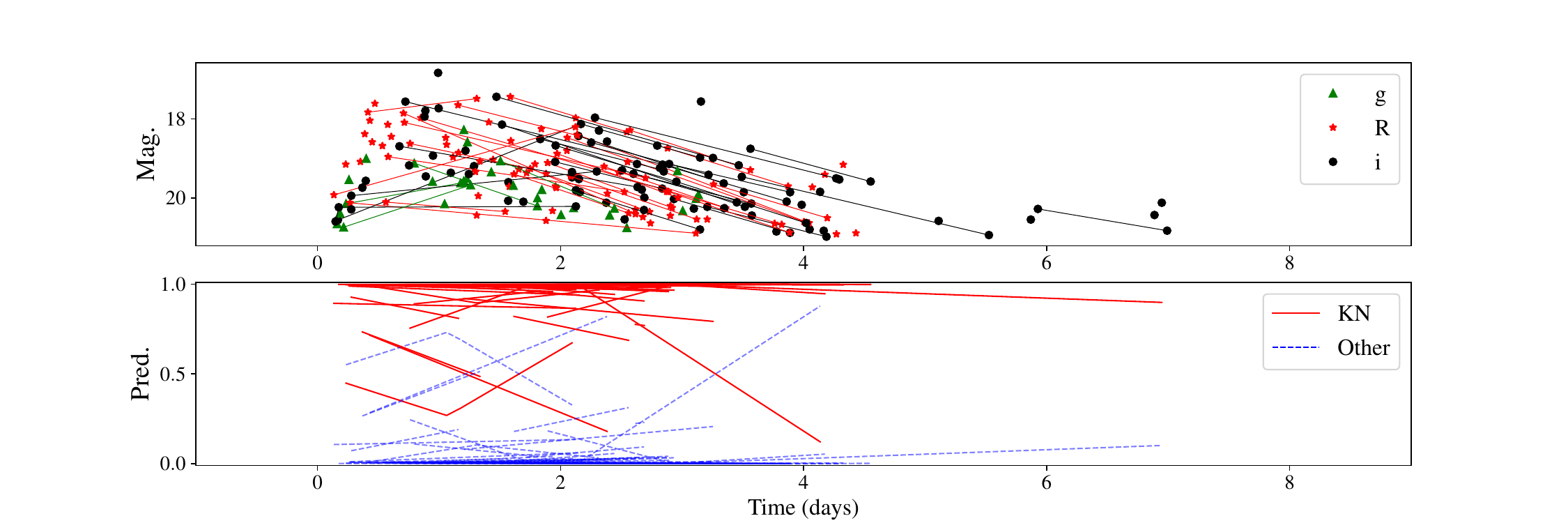}
    \caption{\textbf{Upper panel}: Simulated KNe lightcurves using the
    Rubin Observatory WFD survey cadence. These were not a part of the training sample
    for El-CID. We also generated skymaps for these events using the procedure
    outlined in Sec.~\ref{sec:snana}.
    \textbf{Lower panel}: El-CID predictions on the unseen
    lightcurves. Most of the lightcurves are correctly classified. This
    is a verification that the classifier is learning the temporal features
    and not the cadence of the training set.}
    \label{fig:lsst_result_on_kns}
\end{figure*}
Finally, to verify the performance on a larger sample, and
validate the insensitivity towards cadence, we test the performance of
the mock KN simulated using the Rubin Observatory ``minion\_1016''
cadence for the wide-fast-deep (WFD) survey. This was the previous baseline
cadence adopted by LSST, and was used in
PLAsTiCC~\citep{plasticc, plasticcresults}. We repeat the steps in
Sec.~\ref{sec:training} using the BU19 models to check the performance.
It should be noted that the training set does not contain these simulations.
We still use the O4 noise PSD to generate the skymaps. Therefore, our test models
a hypothetical scenario where Rubin Observatory is in operation during the O4. 

We consider simulations that are within $\sim 100$ Mpc distance. We consider
only the part of the lightcurves where the recovered magnitudes are consistent with 
ZTF's threshold i.e., brighter than 20.5. This results in lightcurves that are around a
week long. As mentioned above, due to the lack of the training set containing KNe that
are observed for larger timescales, the classifier suffers from the caveat that lightcurves
longer than $\sim 1$ week have a tendency to be mis-classified. 

In future work, we plan to make the classifier more robust to instrument thresholds by
including simulations carried out
with different instruments. We overlay $\sim 100$ of these lightcurves in the top panel of 
Fig.~\ref{fig:lsst_result_on_kns}. In the lower panel, we show the performance of the classifier
on these and find that most of the objects are correctly classified -- only 2/115 have their
final classification score as the ``Other'' class, which further validates the performance
on unseen data.

\section{Summary and Conclusion}\label{sec:conclusion}
The third LIGO/Virgo observing run dramatically illustrated the challenge faced by multi-messenger astrophysics -- even as more GW events are identified, locating their EM counterparts quickly becomes daunting because of the large area that needs to be searched, and the large number of contaminating transients that must be filtered out automatically. Additionally, most kilonovae are not expected to be like GW170817/AT~2017gfo, and locating these faint counterparts before they fade into irrelevance demands novel techniques. Here, we have considered a machine learning approach that combines the sparse low S/N optical observations together with contextual information that is already provided with the GW alert, to increase the odds of being able to identify the counterpart. Our results reinforce the utility of deep neural networks in astrophysics, to emulate tasks that previously required visual inspection of the data by expert observers. These techniques will become ever more essential as surveys such as the Rubin Obsevatory discover more sources than can ever be inspected by humans. 

In this article, we have re-purposed the RAPID neural network architecture into El-CID, tuned for
filtering KNe from contaminants. The former has already been shown to
produce promising results, and is used classifying real-time alerts from ZTF as a part of the
ANTARES~\footnote{\url{https://antares.noirlab.edu}} broker~\citep{N18,M21}. While the major challenge towards classifying
fast transients like kilonovae is the sparse data, the use of the other contextual information
available during discovery time has not been made use of in a systematic way i.e., built into
a classification algorithm. This work developed the infrastructure to create realistic end-to-end GW binary merger
simulations, creating a skymap for each, and simulating an associated KN lightcurve in a synoptic
telescope like ZTF.
The procedure we outline allows us to create a rich dataset which emulates the correlations between the
photometric and GW observables. We also simulate the sky -- other objects that can be
contaminants during the search for a counterpart in a GW follow-up. We then supply both the
photometric and the contextual information like the line of sight probability, the angular offset
from the mode of the skymap, and the 90\% confidence sky-localization area as training
feature to help the classifier differentiate between objects even in the absence of good
photometric data. We note that while \cite{Stachie_2020} have used the model trained in MU19
towards classifying KNe, the use of GW discovery data products and its correlations
with photometric information has not been done earlier to our knowledge. Also, the MU19 model is
directed toward SNe classification. For example, in the pre-processing step of MU19, the
lightcurves are interpolated at a 3-day interval, which is sufficient to capture the temporal
features of a SNe, but is large compared to that of a typical KNe. Apart from using a training
set more dense in KNe, we have also adjusted the neural network to shorter timescales.
Although we have restricted to other extra-galactic explosions as contaminants in this work,
the extension to add more contaminant object types, like galactic transients, is
readily done using SNANA which is optimized for high throughput. For example, in this work
the runtime for simulating $\sim 6$ million KN simulations with single node and 20 CPU cores
was $\sim 7$ hours. We, therefore, expect the procedure to scale in terms of adding other
object types. An interesting fast galactic object type in this regard are M-dwarf flares
occurring in the galactic halo. Archival searches for fast optical transients have led to
discovery of several such objects~\citep[for example, see ][]{Ho_2018}. In such a case,
however, other contextual information like the offset from the galactic plane may be used
as another feature during training. This will be considered in the future.
While complete photometric classification into different classes has already been
achieved with RAPID, the key problem we will face during LVK O4 and beyond is to filter out
the KN from the rest of the alert stream from wide-field surveys. We therefore framed El-CID as
a binary classifier, and show that the use of temporal convolutional networks makes the network
largely independent of the observing cadence. We achieve a $\sim 99\%$ accuracy with our training
sample, and also verify the performance of the classifier on unseen data.
We find that the
classifier works expectedly for the only EMGW event, GW170817/AT~2017gfo, and the classification
is correct across the data taken by different instruments. We also analyze the case of AT~2019npv
which was a contaminant in the follow-up effort of GW190814, and find that the classifier is
able to rule out the object in the first few epochs of observation.

Tools like El-CID will be crucial in the follow-up efforts in the future. We would like
to stress that the use case is for broker teams to slice through the candidate list in near real
time after discovery. We would also like to point out that the values given by El-CID is to be
interpreted as a \emph{score}, and is not analogous to an evidence from a bayesian inference.
Retrospective analysis on a population should use the values from El-CID accordingly.
We expect the El-CID architecture can be periodically trained from nightly data taken from
public brokers like ANTARES during Rubin Observatory operations. Dedicated public EMGW efforts like the Scalable Cyberinfrastructure
for Multi-messenger Astrophysics (SCiMMA)~\footnote{\url{https://scimma.org/}} could have El-CID running
as an annotation service. The results provided by El-CID could be used by downstream facilities to
schedule observations. Recently, there have been conceptually similar efforts like the ZTFRest
framework~\citep{andreoni2021fasttransient} that provides a service for automated kilonova
searches. Another example of an ``add-on'' to existing facilities is the Gamma-ray Urgent
Archiver for Novel Opportunities (GUANO) toolkit~\citep{Tohuvavohu_2020} that performs
low-latency searches for subthreshold/off-axis GRBs and downlinks interesting candidates for
deeper searches which would otherwise be lost from traditional search algorithms.
Given the rarity of KNe, no individual approach at this stage can serve as the perfect solution and
therefore having complementary efforts will be crucial for the future of multi-messenger astrophysics.

\section*{Acknowledgements}
D.~C. is supported by the Illinois Survey Science Fellowship from the Center for
AstroPhysical Surveys (CAPS)~\footnote{\url{https://caps.ncsa.illinois.edu/}}
at the National Center for Supercomputing Applications (NCSA), University of Illinois
Urbana-Champaign. D.~C. acknowledges computing resources provided by CAPS to carry our
this research. 
This work made use of the Illinois Campus Cluster, a computing resource that is operated
by the Illinois Campus Cluster Program (ICCP) in conjunction with NCSA which is supported
by funds from the University of Illinois at Urbana-Champaign.
This research also used resources of the National Energy Research Scientific Computing
Center (NERSC), a U.S. Department of Energy Office of Science User Facility located at
Lawrence Berkeley National Laboratory, operated under Contract No. DE-AC02-05CH11231.

D.~C. is grateful to Richard Kessler for help with using SNANA. D.~C. would like to
thank Igor Andreoni for helpful discussions, Alex Gagliano for comments on the draft,
Michael Coughlin and Mattia Bulla for help with the BU19 models.
The authors specially thank Leo Singer for critical comments. The authors
would also like to thank the anonymous referee for helpful comments.

\section*{Data Availability}
The SED models used in the study are a part of SNANA~\citep{snana} package data
available at \url{https://zenodo.org/record/4001178#.YR3qAFtOlcA}. The skymaps
were generated by running the BAYESTAR~\citep{singer_and_price} code. The neural
network architecture used Tensorflow~\citep{tensorflow2015-whitepaper} and
Keras~\citep{keras}. Currently, the neural network architecture is implemented as a 
branch of \texttt{astrorapid} in \url{https://github.com/deepchatterjeeligo/astrorapid/tree/kn-rapid}.
Eventually, this will be moved to a new public repository. Codes for producing
lightcurves from representative SED models in Fig.~\ref{fig:gw170817_lightcurve_fits}
are avaiable in \url{https://github.com/deepchatterjeeligo/bulla-models}.
A prototype trained classifier is demonstrated at
\url{https://github.com/broker-workshop/tutorials/blob/main/ANTARES/lsst-broker-workshop-mma-demo.ipynb}.
Trained weights and training data for this work will be made available upon request to
the authors.

\appendix
\section{Estimating detection efficiency for ZTF}\label{appendix:efficiency_ztf}
\begin{figure*}
    \centering
    \includegraphics[width=1.0\linewidth]{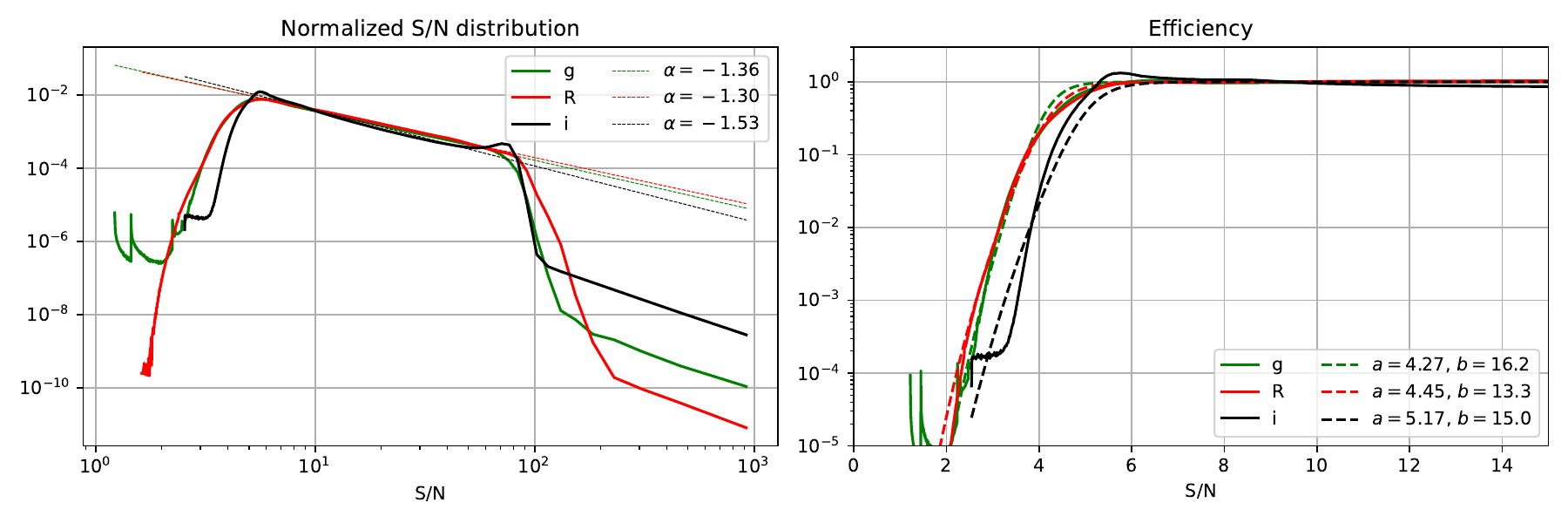}
    \caption{Signal-to-noise distributions. \textbf{Left panel}: The distribution of ZTF DR3 detections over $S/N$ normalized to have have unity integral (solid lines). Thin dashed lines show power-law fits for these distributions for the interval $6 < S/N < 80$, corresponding exponents $\alpha$ are listed in the legend. \textbf{Right panel}: The detection efficiency distribution. Solid lines show the ratio of $S/N$ distributions to corresponding fits (solid to dashed lines from the left panel). Dashed lines show $(1 + (a/(S/N))^b)^{-1}$ fits of these distributions, fit parameters are given in the legend.}
    \label{fig:eff}
\end{figure*}

Here, we show how we estimated the detection efficiency of ZTF: the probability of
a source detection for given flux signal-to-noise ratio $\mathrm{eff}(S/N)$.
We used the magnitude uncertainty $\sigma_m$ from the third data release of ZTF (ZTF DR3)\footnote{\url{http://ztf.caltech.edu/page/dr3}} to plot distribution of the
number of defections $n(S/N)$ assuming that $S/N = \ln{\left(10^{0.4}\right)} / \sigma_m$.
This distribution can be considered as $n(S/N) = n_\mathrm{tot}\, p_\mathrm{S/N}(S/N)\, \mathrm{eff}(S/N)$ where $n_\mathrm{tot}$ is the total number of occurred events, and
$p_\mathrm{S/N}(S/N)$ is a probability density function of all (detected and non-detected) events.
We assumed that for $10 \lesssim S/N \lesssim 100$
(it roughly corresponds to $21 \gtrsim m \gtrsim 14.5$ for $R$ passband and under an assumption of the Poisson noise)
the detection is 100\% effective
which means that for this interval $n(S/N) \sim p_\mathrm{S/N}(S/N)$. We found that for
considering $S/N$ interval the distribution can be fitted by power-law function 
$\left[\beta\,(S/N)\right]^\alpha$ with exponent $\alpha=-1.36$ for $g$, $-1.30$ for $R$,
and $-1.53$ for $i$. Note that we would have $\alpha=-3/2$ for Poisson noise and isotropic
distribution of sources of the same luminosity, which is close to found values and
allows us to consider that $p_\mathrm{S/N}(S/N) \sim (S/N)^\alpha$ can be used for
all $S/N$ values.
So under this framework $\mathrm{eff} = n(S/N) / \left[\beta\,(S/N)\right]^\alpha$.
For simplicity and to avoid efficiency to be greater 100\% we fitted obtained efficiency curve by a sigmoid function $(1 + (a/(S/N))^b)^{-1}$.
In this work we aren't interested in detection efficiency for
bright objects, therefore we ignore $S/N \gg 10$ when performed the last fit.
In the left panel of Fig.~\ref{fig:eff} we plot normalized $S/N$ distributions and their power-law fits, in its right panel we plot efficiency distribution and their sigmoid fits.
Finally, we multiplied $(S/N)$ by the factor of $\sqrt{2}$ to take into
account the noise level difference in PSF photometry of ZTF DR3 and 
differential photometry of the ZTF alert stream under of assumption that image
subtraction leads to twice large variance of the source flux estimate.

\section{Example Lightcurves}\label{appendix:example_lightcurves}
\begin{figure*}
    \centering
    \includegraphics[width=0.32\textwidth]{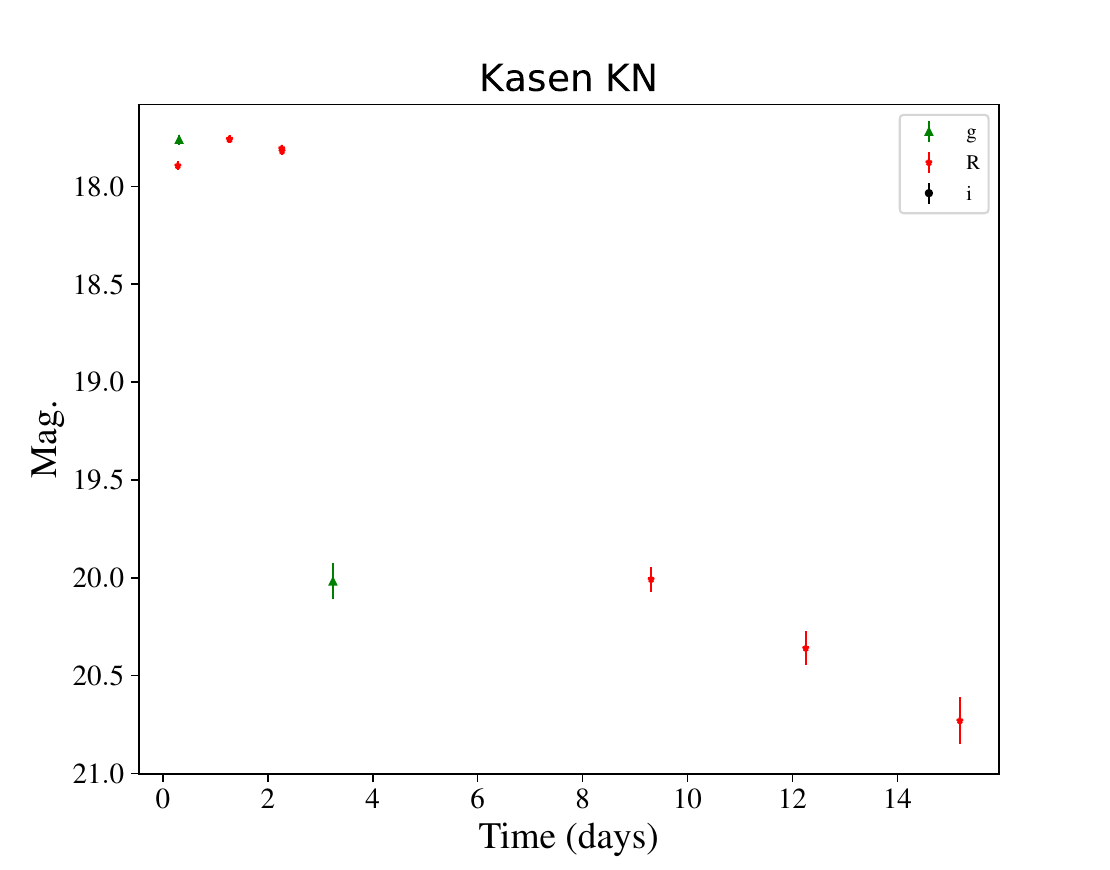}
    \includegraphics[width=0.32\textwidth]{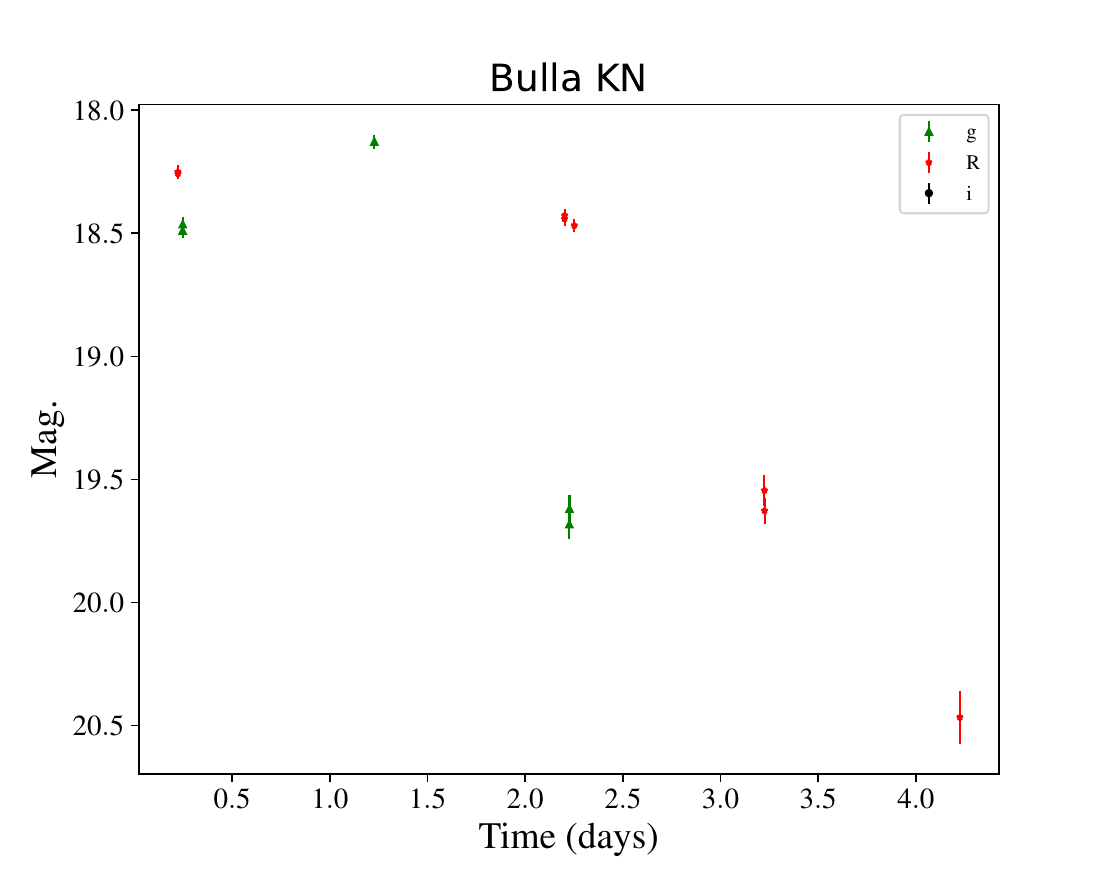}
    \includegraphics[width=0.32\textwidth]{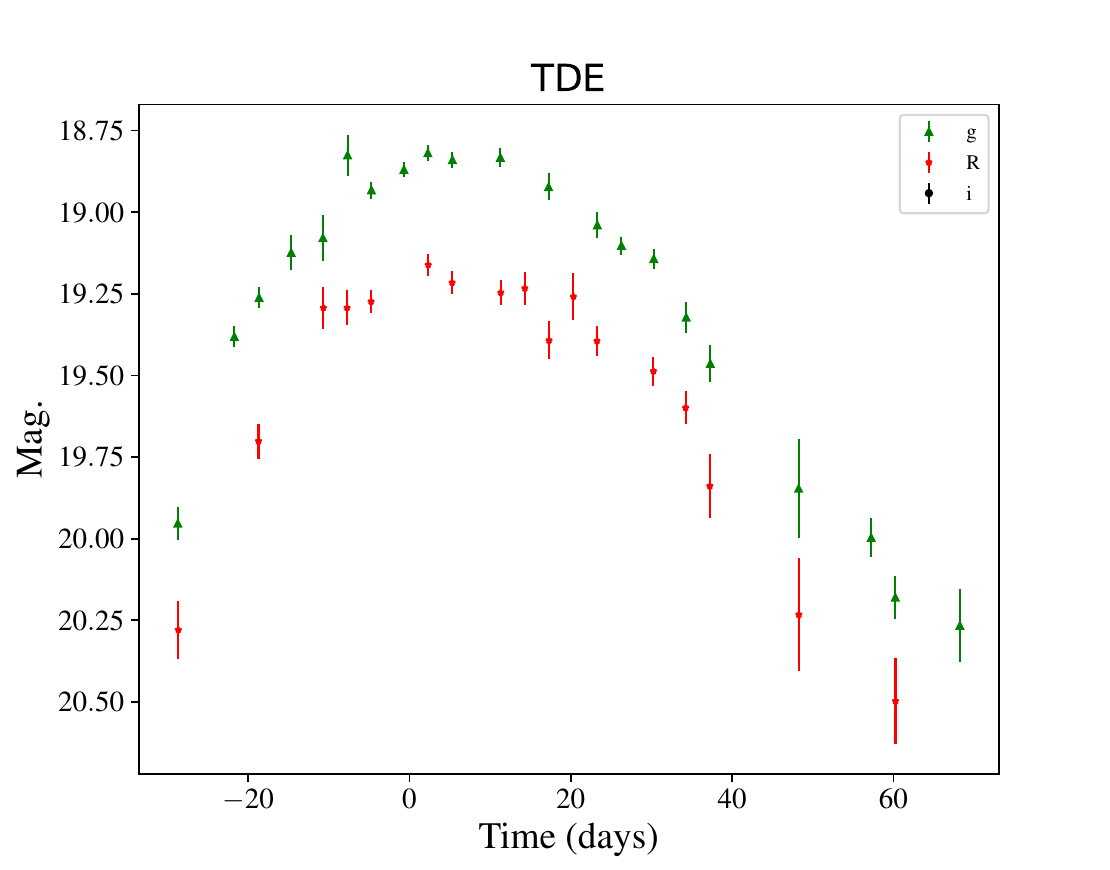}
    \\
    \includegraphics[width=0.32\textwidth]{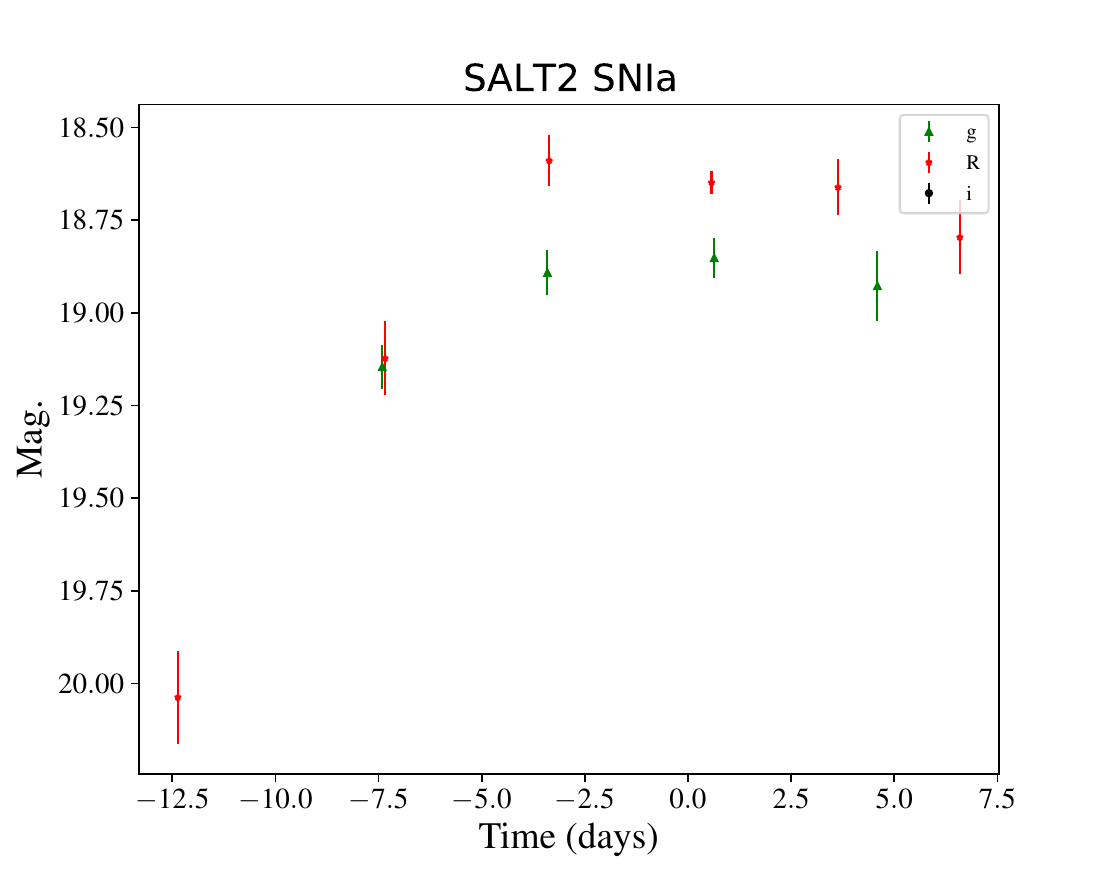}
    \includegraphics[width=0.32\textwidth]{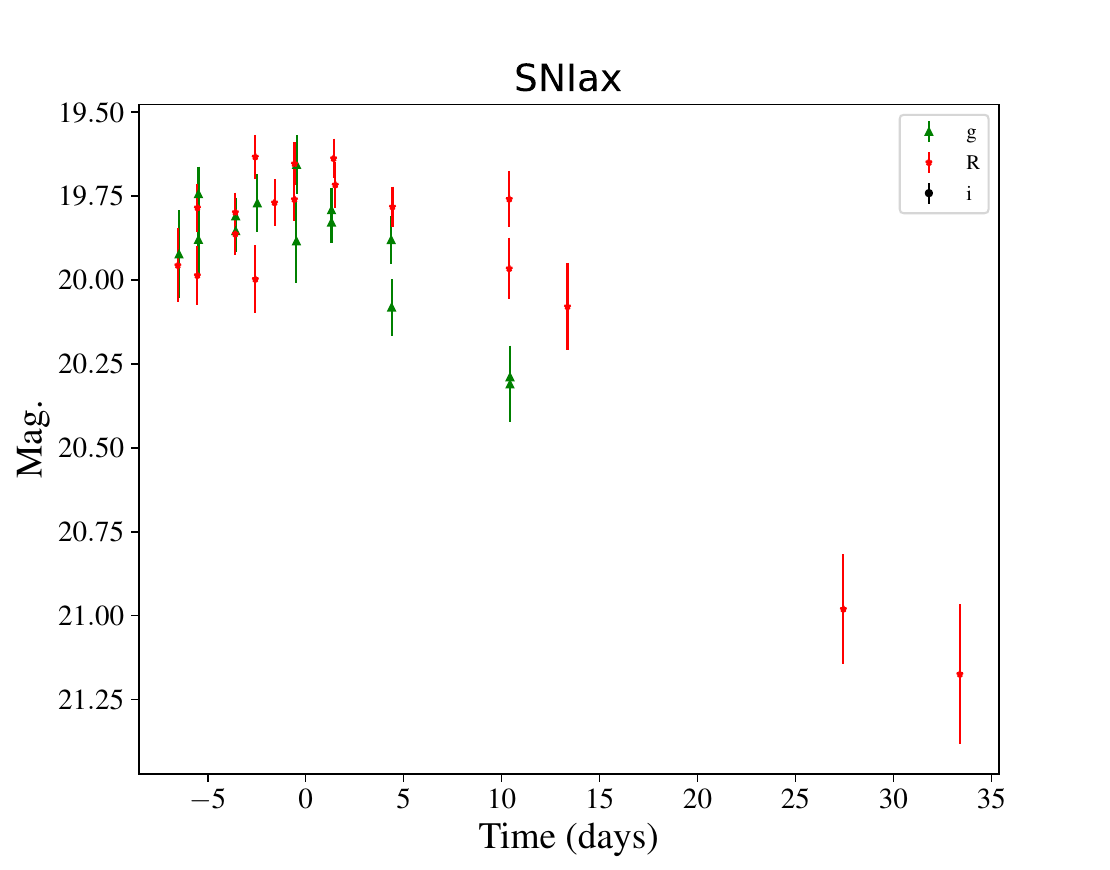}
    \includegraphics[width=0.32\textwidth]{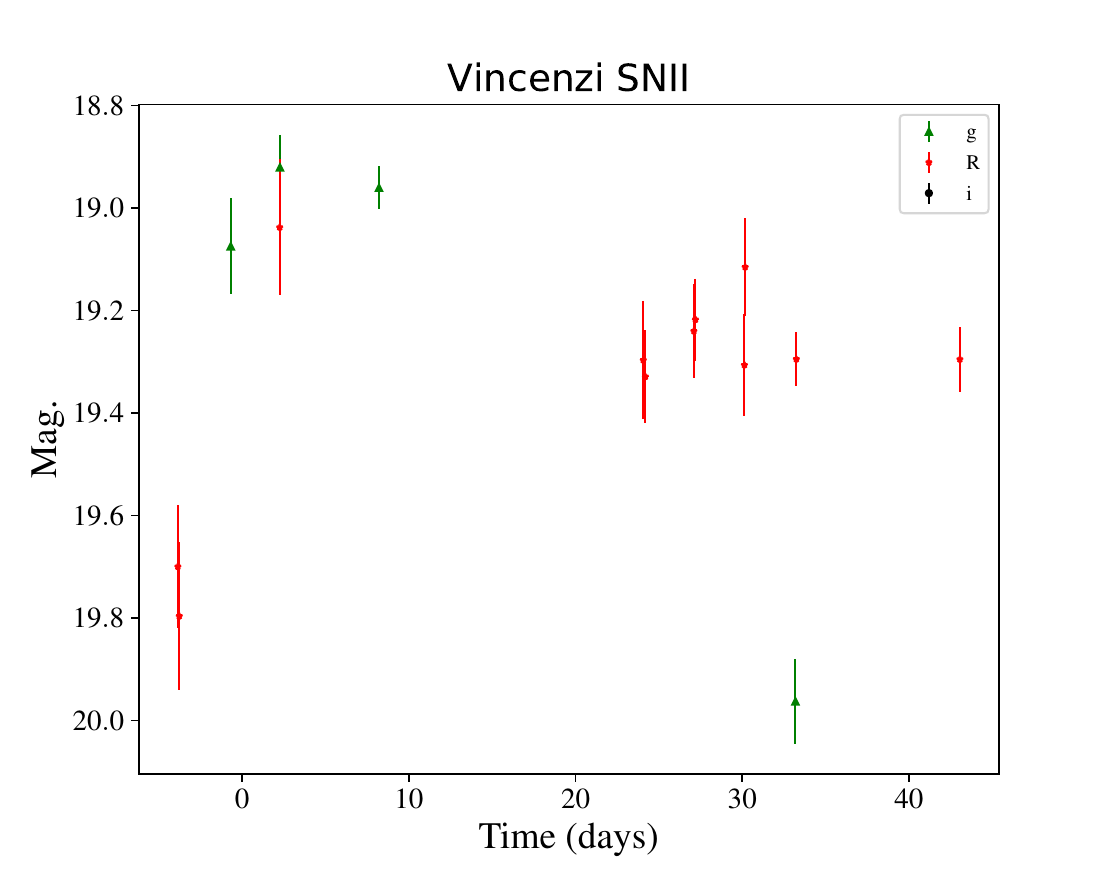}
    \\
    \includegraphics[width=0.32\textwidth]{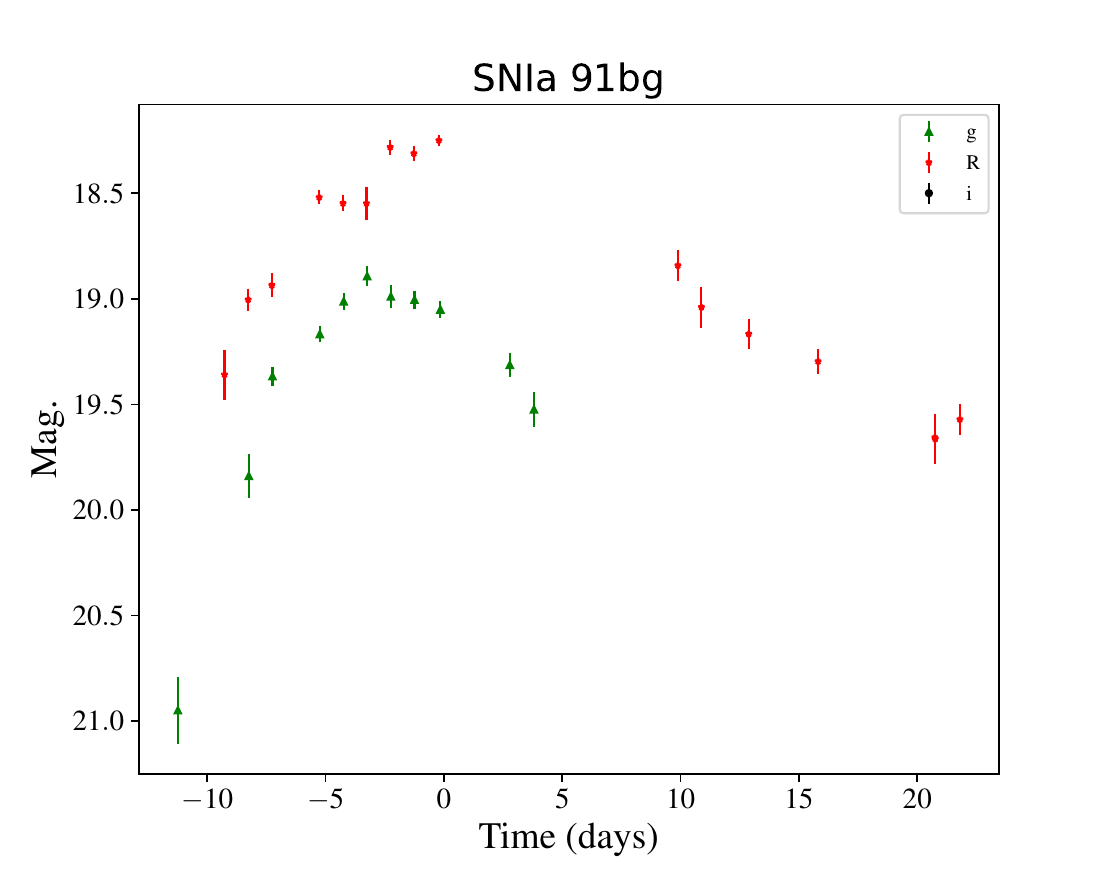}
    \includegraphics[width=0.32\textwidth]{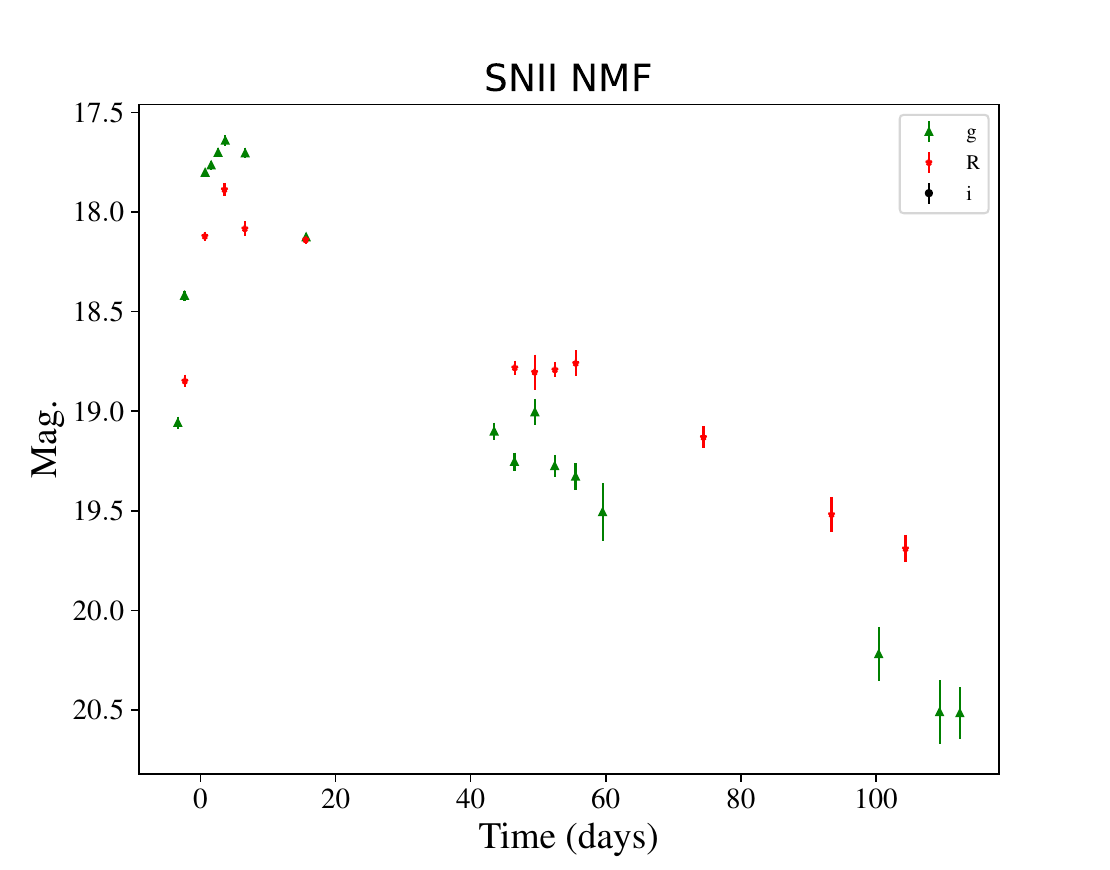}
    \includegraphics[width=0.32\textwidth]{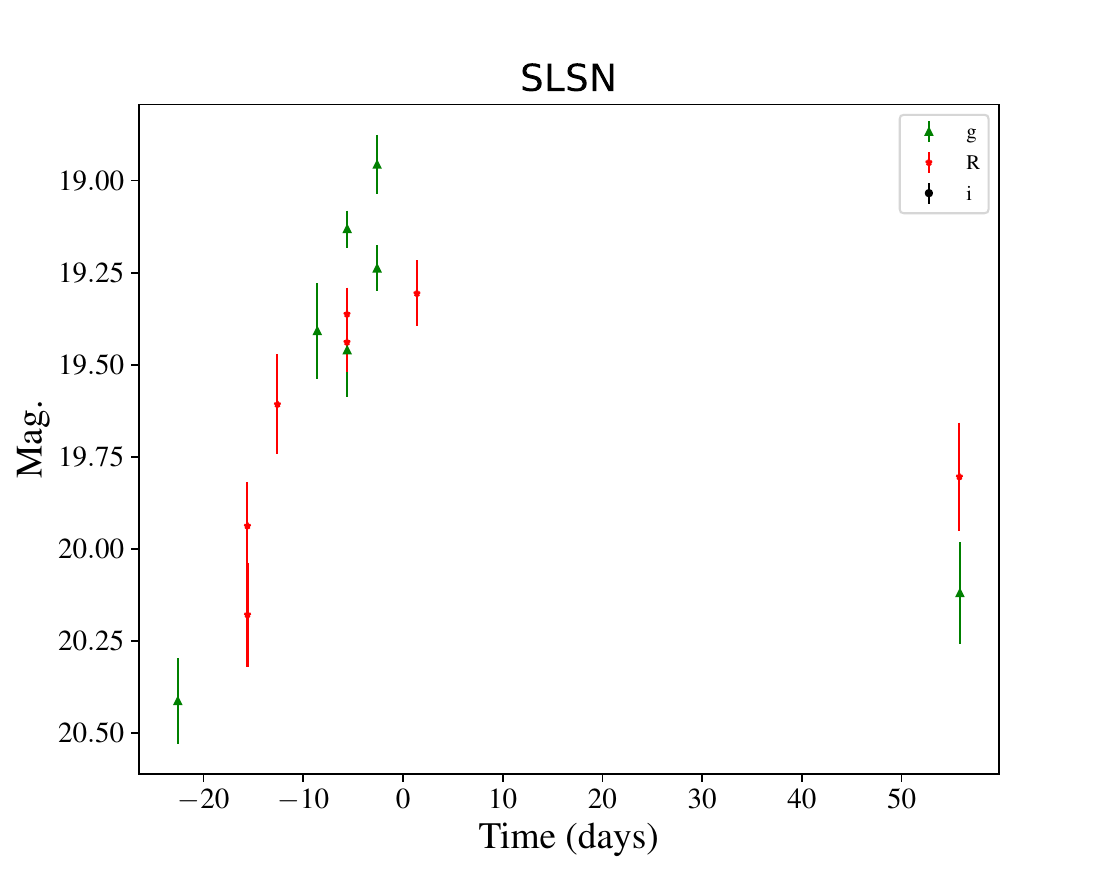}
    \caption{A random sample of example lightcurves from the training set}
    \label{fig:appendix_example_lightcurves}
\end{figure*}
In Fig.~\ref{fig:appendix_example_lightcurves} we show a few randomly chosen example
lightcurves from the training set used in this work.

\section{Variation of {\mej} with EOS}\label{appendix:mej_eos_variation}
\begin{figure}
    \centering
    \includegraphics[width=1.0\columnwidth]{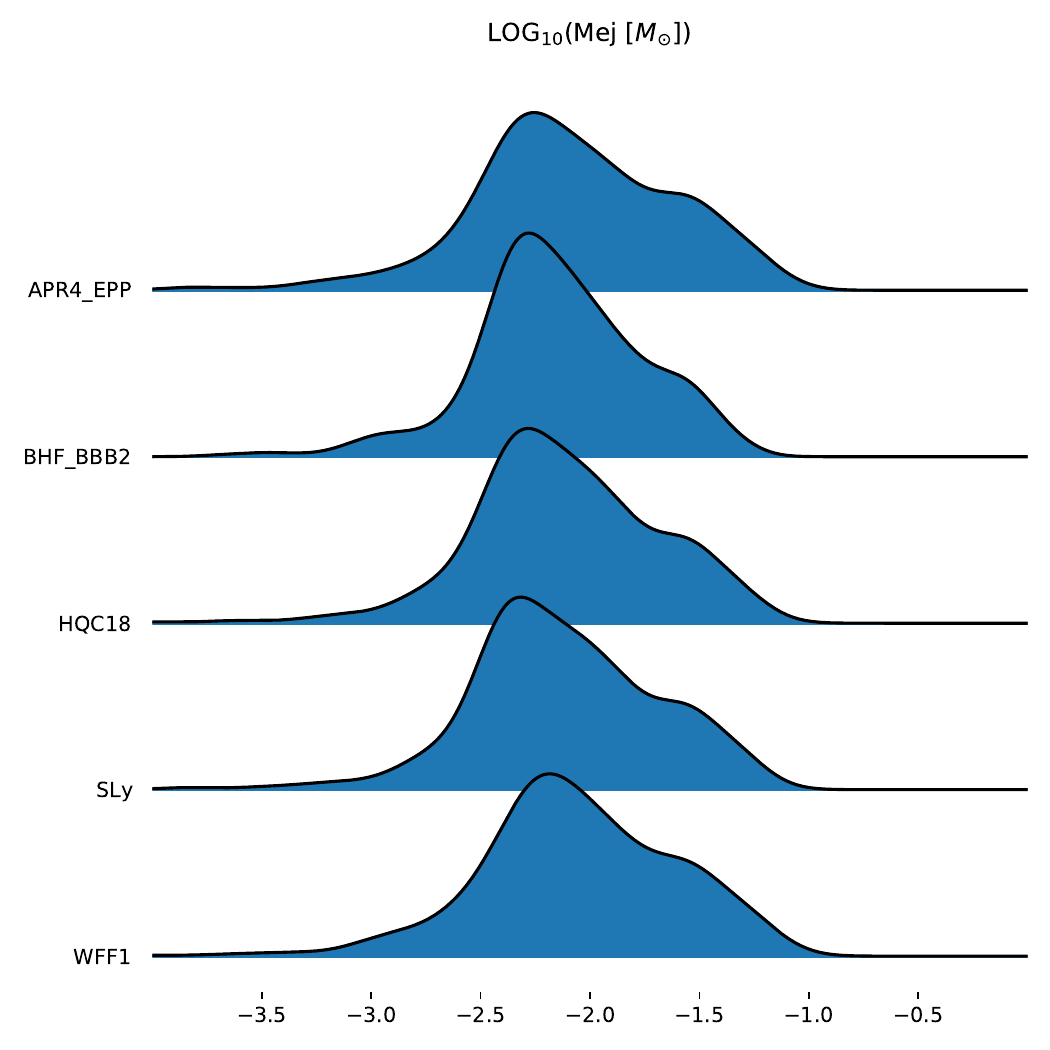}
    \caption{This figure shows the distribution of the variation of ejecta mass,
    {\mej}, with EoS.}
    \label{fig:mej_distribution_eos}
\end{figure}
The ejecta mass in Eq.~(\ref{eq:ejecta_mass_fit}) can be a strong function of the stiffness
of the NS. Stiffer stars allow for larger radii at the same mass, and therefore are easily
tidally disrupted. However, the tidal effects from GW170817 constrains several of the
EoSs in the literature. Recently, \cite{2021arXiv210408681G} have performed model
selection of representative EoSs against tidal deformability measurements of GW170817.
In Fig.~\ref{fig:mej_distribution_eos}, we restrict to those that have the highest evidences,
and assess the variation of the ejecta mass from the DU17 relation. We sample, component
masses uniformly $\in [1, 3]\;M_{\odot}$, rejecting combinations in case they are greater
than the maximum mass allowed by an EoS. We find that the distribution of the ejecta
mass is not significantly affected in this case. Hence, our choice of using APR4\_EPP
is justified. The model grids used in this work have an spacing of $\mathcal{O}(10^{-2})$
in ejecta mass which is large compared to the differences in ejecta masses from the
strongly supported EoSs.

\section{Results from LVK design sensitivity}\label{appendix:design_sensitivity}
\begin{figure*}
    \centering
    \includegraphics[width=0.32\textwidth]{DECam.pdf}
    \includegraphics[width=0.32\textwidth]{Magellan.pdf}
    \includegraphics[width=0.32\textwidth]{Swope.pdf}
    \\
    \includegraphics[width=0.32\textwidth]{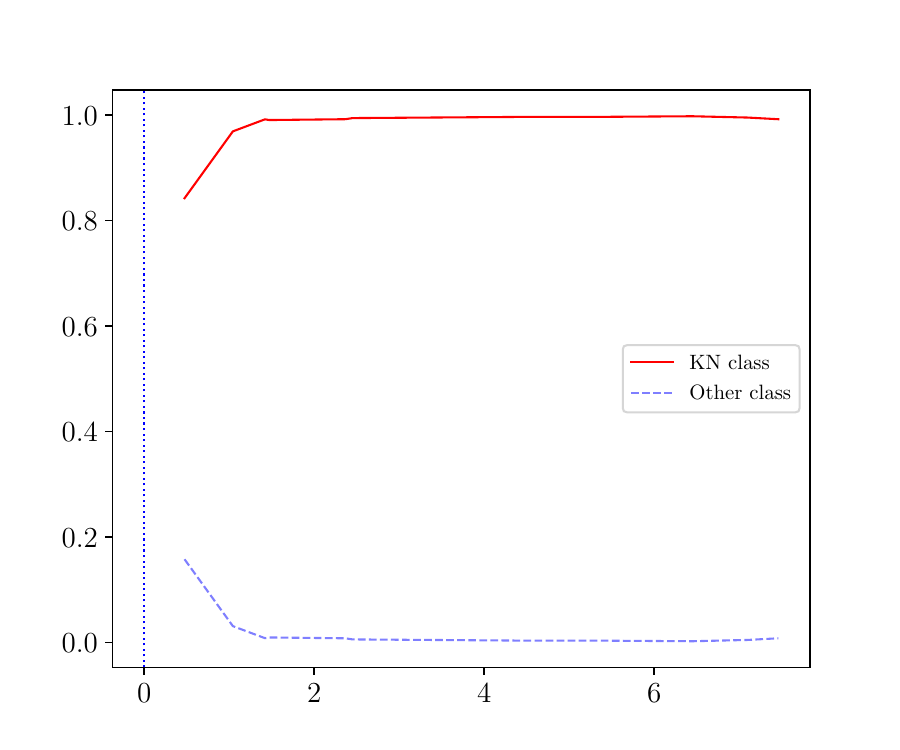}
    \includegraphics[width=0.32\textwidth]{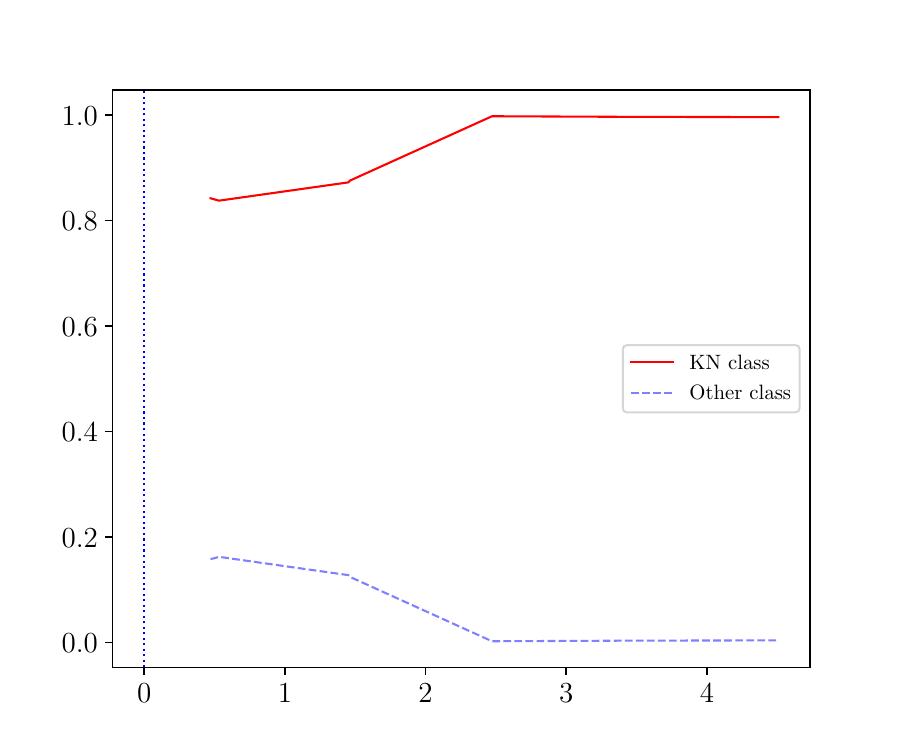}
    \includegraphics[width=0.32\textwidth]{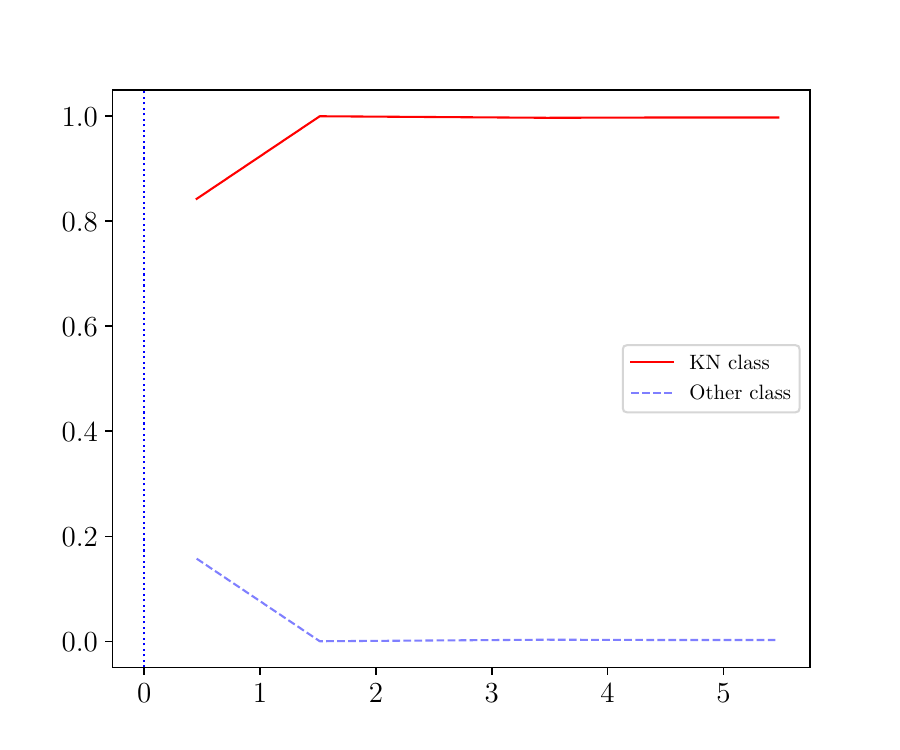}
    \\
    \includegraphics[width=0.32\textwidth]{LCO-1m.pdf}
    \includegraphics[width=0.32\textwidth]{PS1.pdf}
    \includegraphics[width=0.32\textwidth]{HST.pdf}
    \\
    \includegraphics[width=0.32\textwidth]{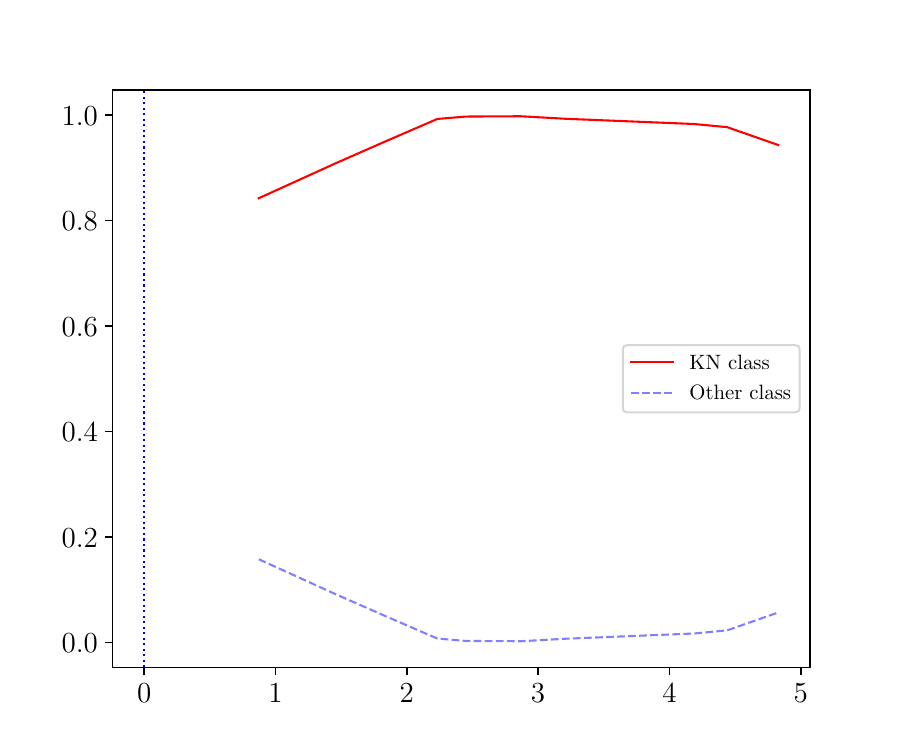}
    \includegraphics[width=0.32\textwidth]{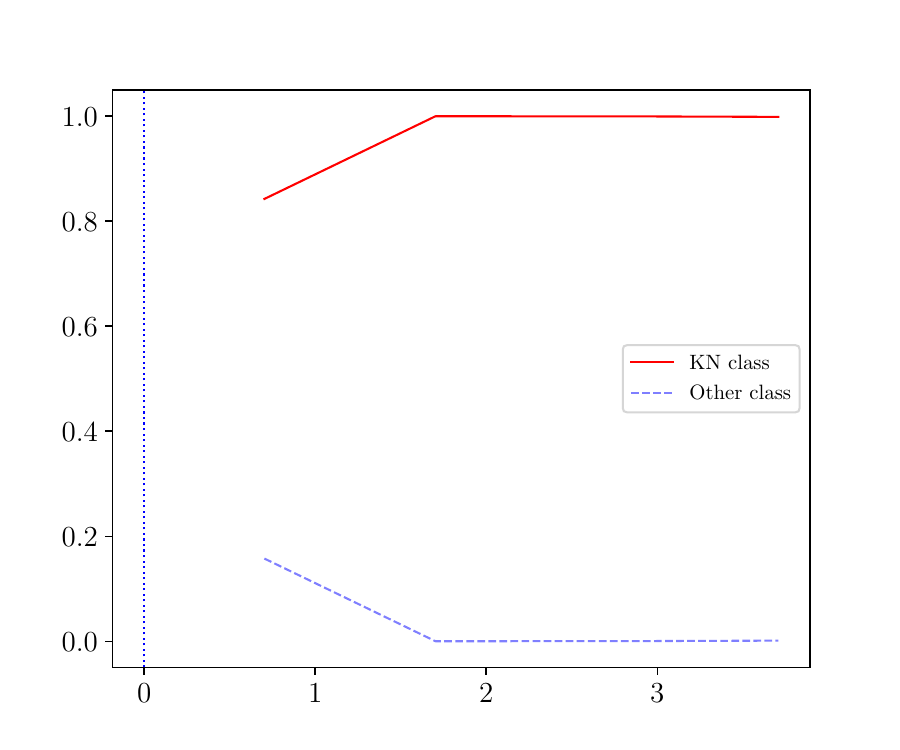}
    \includegraphics[width=0.32\textwidth]{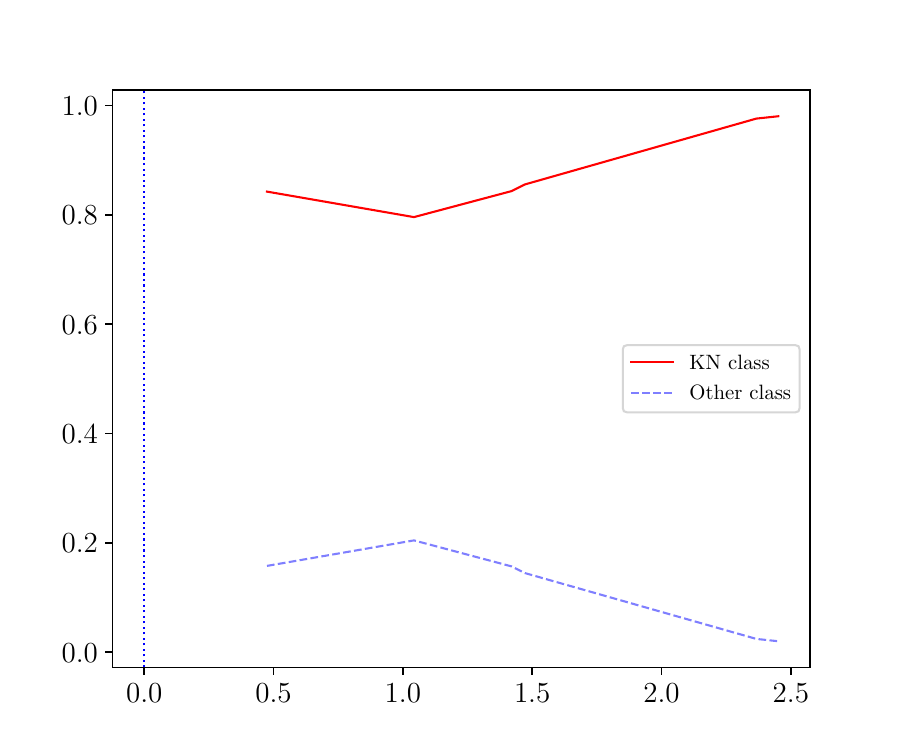}
    \caption{Similar to Fig.~\ref{fig:predictions_GW170817} except the
    the classifier trained on design sensitivity era
    on $\sim 1$ week data for GW170817
    obtained from the Open Kilonova catalog.}
    \label{fig:appendix_predictions_GW170817}
\end{figure*}

\begin{figure}
    \centering
    \includegraphics[width=1.0\columnwidth, trim=0cm 0cm 0cm 1cm, clip]
        {DG19wxnjc.pdf}\\
    \includegraphics[width=1.0\columnwidth, trim=0cm 0cm 0cm 1cm, clip]
        {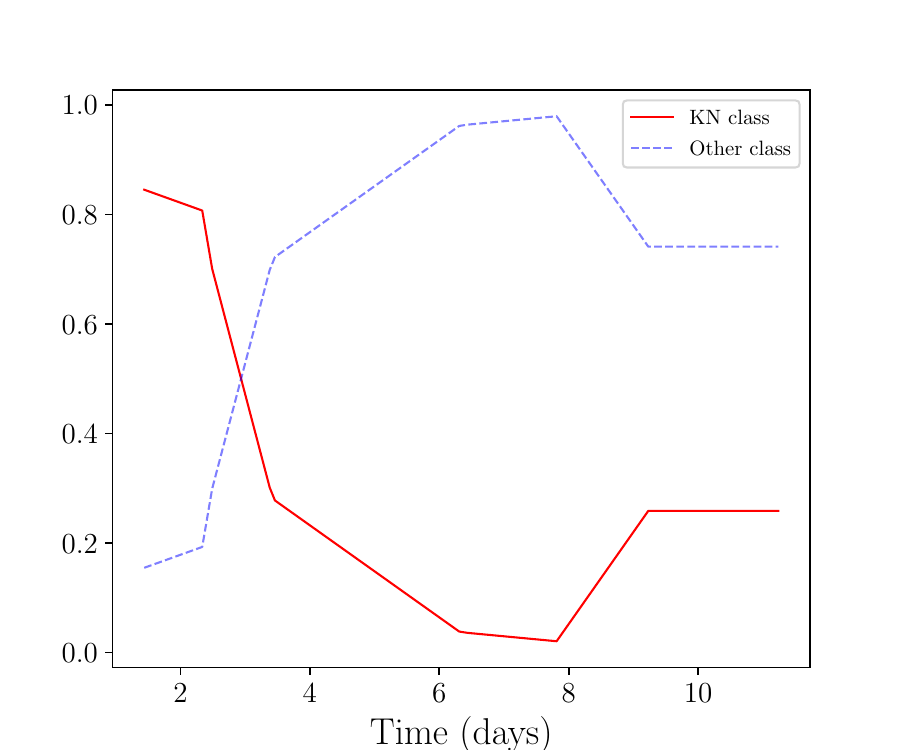}
    \caption{AT~2019npv classification similar to Fig.~\ref{fig:at2019_npv},
    except the classifier is trained on simulations in the design sensitivity era.
    We find that the results are robust irrespective of the noise-curve
    between O4 and design sensitivity.}
    \label{fig:appendix_at2019_npv}
\end{figure}
In this section, we repeat the training using a dataset set where we use the same objects
but consider a design-sensitivity noise-curve for LIGO/Virgo/KAGRA. We also
restrict to only events that are detected in at least 3-detectors.
In Fig.~\ref{fig:appendix_predictions_GW170817} and Fig.~\ref{fig:appendix_at2019_npv},
we show the performance of this classifier on AT~2017gfo and AT~2019npv
as in Fig.~\ref{fig:predictions_GW170817} and Fig.~\ref{fig:at2019_npv} in the main
body of the text. We find that the results are not strongly sensitive to the change in
detector sensitivity.

\bibliographystyle{mnras}
\bibliography{references}

\begin{thebibliography}{}
\makeatletter
\relax
\def\mn@urlcharsother{\let\do\@makeother \do\$\do\&\do\#\do\^\do\_\do\%\do\~}
\def\mn@doi{\begingroup\mn@urlcharsother \@ifnextchar [ {\mn@doi@}
  {\mn@doi@[]}}
\def\mn@doi@[#1]#2{\def\@tempa{#1}\ifx\@tempa\@empty \href
  {http://dx.doi.org/#2} {doi:#2}\else \href {http://dx.doi.org/#2} {#1}\fi
  \endgroup}
\def\mn@eprint#1#2{\mn@eprint@#1:#2::\@nil}
\def\mn@eprint@arXiv#1{\href {http://arxiv.org/abs/#1} {{\tt arXiv:#1}}}
\def\mn@eprint@dblp#1{\href {http://dblp.uni-trier.de/rec/bibtex/#1.xml}
  {dblp:#1}}
\def\mn@eprint@#1:#2:#3:#4\@nil{\def\@tempa {#1}\def\@tempb {#2}\def\@tempc
  {#3}\ifx \@tempc \@empty \let \@tempc \@tempb \let \@tempb \@tempa \fi \ifx
  \@tempb \@empty \def\@tempb {arXiv}\fi \@ifundefined
  {mn@eprint@\@tempb}{\@tempb:\@tempc}{\expandafter \expandafter \csname
  mn@eprint@\@tempb\endcsname \expandafter{\@tempc}}}

\bibitem[\protect\citeauthoryear{Abadi et~al.,}{Abadi
  et~al.}{2015}]{tensorflow2015-whitepaper}
Abadi M.,  et~al., 2015, {TensorFlow}: Large-Scale Machine Learning on
  Heterogeneous Systems, \url {https://www.tensorflow.org/}

\bibitem[\protect\citeauthoryear{Abbott et~al.,}{Abbott
  et~al.}{2016}]{Abbott_2016}
Abbott B.,  et~al., 2016, \mn@doi [Physical Review Letters]
  {10.1103/physrevlett.116.061102}, 116

\bibitem[\protect\citeauthoryear{Abbott et~al.,}{Abbott
  et~al.}{2017a}]{gw170817}
Abbott B.,  et~al., 2017a, \mn@doi [Physical Review Letters]
  {10.1103/physrevlett.119.161101}, 119

\bibitem[\protect\citeauthoryear{{Abbott} et~al.,}{{Abbott}
  et~al.}{2017b}]{2017Natur.551...85A}
{Abbott} B.~P.,  et~al., 2017b, \mn@doi [\nat] {10.1038/nature24471}, \href
  {https://ui.adsabs.harvard.edu/abs/2017Natur.551...85A} {551, 85}

\bibitem[\protect\citeauthoryear{{Abbott} et~al.,}{{Abbott}
  et~al.}{2017c}]{mma_2017}
{Abbott} B.~P.,  et~al., 2017c, \mn@doi [ApJ Lett.] {10.3847/2041-8213/aa91c9},
  \href {http://adsabs.harvard.edu/abs/2017ApJ...848L..12A} {848, L12}

\bibitem[\protect\citeauthoryear{{Abbott} et~al.,}{{Abbott}
  et~al.}{2018a}]{2018LRR....21....3A}
{Abbott} B.~P.,  et~al., 2018a, \mn@doi [Living Reviews in Relativity]
  {10.1007/s41114-018-0012-9}, \href
  {https://ui.adsabs.harvard.edu/abs/2018LRR....21....3A} {21, 3}

\bibitem[\protect\citeauthoryear{{Abbott} et~al.,}{{Abbott}
  et~al.}{2018b}]{2018PhRvL.121p1101A}
{Abbott} B.~P.,  et~al., 2018b, \mn@doi [\prl]
  {10.1103/PhysRevLett.121.161101}, \href
  {https://ui.adsabs.harvard.edu/abs/2018PhRvL.121p1101A} {121, 161101}

\bibitem[\protect\citeauthoryear{{Abbott} et~al.,}{{Abbott}
  et~al.}{2020a}]{2020arXiv201014527A}
{Abbott} R.,  et~al., 2020a, arXiv e-prints, \href
  {https://ui.adsabs.harvard.edu/abs/2020arXiv201014527A} {p. arXiv:2010.14527}

\bibitem[\protect\citeauthoryear{Abbott et~al.,}{Abbott
  et~al.}{2020b}]{Abbott_2020}
Abbott R.,  et~al., 2020b, \mn@doi [The Astrophysical Journal]
  {10.3847/2041-8213/ab960f}, 896, L44

\bibitem[\protect\citeauthoryear{Andreoni et~al.,}{Andreoni
  et~al.}{2020}]{Andreoni_2020}
Andreoni I.,  et~al., 2020, \mn@doi [The Astrophysical Journal]
  {10.3847/1538-4357/ab6a1b}, 890, 131

\bibitem[\protect\citeauthoryear{Andreoni et~al.,}{Andreoni
  et~al.}{2021}]{andreoni2021fasttransient}
Andreoni I.,  et~al., 2021, Fast-transient Searches in Real Time with ZTFReST:
  Identification of Three Optically-discovered Gamma-ray Burst Afterglows and
  New Constraints on the Kilonova Rate (\mn@eprint {arXiv} {2104.06352})

\bibitem[\protect\citeauthoryear{Antier et~al.,}{Antier
  et~al.}{2019}]{Antier_2019}
Antier S.,  et~al., 2019, \mn@doi [Monthly Notices of the Royal Astronomical
  Society] {10.1093/mnras/stz3142}, 492, 3904–3927

\bibitem[\protect\citeauthoryear{{Arcavi} et~al.,}{{Arcavi}
  et~al.}{2017}]{arcavi_2017}
{Arcavi} I.,  et~al., 2017, \mn@doi [Nature] {10.1038/nature24291}, \href
  {https://ui.adsabs.harvard.edu/abs/2017Natur.551...64A} {551, 64}

\bibitem[\protect\citeauthoryear{Arun, Iyer, Sathyaprakash  \&
  Sundararajan}{Arun et~al.}{2005}]{PhysRevD.71.084008}
Arun K.~G.,  Iyer B.~R.,  Sathyaprakash B.~S.,   Sundararajan P.~A.,  2005,
  \mn@doi [Phys. Rev. D] {10.1103/PhysRevD.71.084008}, 71, 084008

\bibitem[\protect\citeauthoryear{{Bai}, {Zico Kolter}  \& {Koltun}}{{Bai}
  et~al.}{2018}]{tcn}
{Bai} S.,  {Zico Kolter} J.,   {Koltun} V.,  2018, arXiv e-prints, \href
  {https://ui.adsabs.harvard.edu/abs/2018arXiv180301271B} {p. arXiv:1803.01271}

\bibitem[\protect\citeauthoryear{{Bellm} et~al.,}{{Bellm}
  et~al.}{2019}]{2019PASP..131a8002B}
{Bellm} E.~C.,  et~al., 2019, \mn@doi [\pasp] {10.1088/1538-3873/aaecbe}, \href
  {https://ui.adsabs.harvard.edu/abs/2019PASP..131a8002B} {131, 018002}

\bibitem[\protect\citeauthoryear{{Blinnikov}, {Novikov}, {Perevodchikova}  \&
  {Polnarev}}{{Blinnikov} et~al.}{1984}]{1984SvAL...10..177B}
{Blinnikov} S.~I.,  {Novikov} I.~D.,  {Perevodchikova} T.~V.,   {Polnarev}
  A.~G.,  1984, Soviet Astronomy Letters, \href
  {https://ui.adsabs.harvard.edu/abs/1984SvAL...10..177B} {10, 177}

\bibitem[\protect\citeauthoryear{{Bulla}}{{Bulla}}{2019}]{bulla_2019}
{Bulla} M.,  2019, \mn@doi [\mnras] {10.1093/mnras/stz2495}, \href
  {https://ui.adsabs.harvard.edu/abs/2019MNRAS.489.5037B} {489, 5037}

\bibitem[\protect\citeauthoryear{Chatterjee, Nugent, Brady, Cannella, Kaplan
  \& Kasliwal}{Chatterjee et~al.}{2019}]{Chatterjee_2019}
Chatterjee D.,  Nugent P.~E.,  Brady P.~R.,  Cannella C.,  Kaplan D.~L.,
  Kasliwal M.~M.,  2019, \mn@doi [The Astrophysical Journal]
  {10.3847/1538-4357/ab2b9c}, 881, 128

\bibitem[\protect\citeauthoryear{Chollet et~al.}{Chollet et~al.}{2015}]{keras}
Chollet F.,  et~al., 2015, Keras, \url{https://keras.io}

\bibitem[\protect\citeauthoryear{{Coughlin}}{{Coughlin}}{2020}]{2020NatAs...4..550C}
{Coughlin} M.~W.,  2020, \mn@doi [Nature Astronomy]
  {10.1038/s41550-020-1130-3}, \href
  {https://ui.adsabs.harvard.edu/abs/2020NatAs...4..550C} {4, 550}

\bibitem[\protect\citeauthoryear{{Coughlin} et~al.,}{{Coughlin}
  et~al.}{2020}]{coughlin_2020}
{Coughlin} M.~W.,  et~al., 2020, \mn@doi [Nature Communications]
  {10.1038/s41467-020-17998-5}, \href
  {https://ui.adsabs.harvard.edu/abs/2020NatCo..11.4129C} {11, 4129}

\bibitem[\protect\citeauthoryear{Coulter et~al.,}{Coulter
  et~al.}{2017}]{Coulter_2017}
Coulter D.~A.,  et~al., 2017, \mn@doi [Science] {10.1126/science.aap9811}, 358,
  1556

\bibitem[\protect\citeauthoryear{{Dietrich} \& {Ujevic}}{{Dietrich} \&
  {Ujevic}}{2017}]{2017CQGra..34j5014D}
{Dietrich} T.,  {Ujevic} M.,  2017, \mn@doi [Classical and Quantum Gravity]
  {10.1088/1361-6382/aa6bb0}, \href
  {https://ui.adsabs.harvard.edu/abs/2017CQGra..34j5014D} {34, 105014}

\bibitem[\protect\citeauthoryear{Dobie et~al.,}{Dobie
  et~al.}{2019}]{Dobie_2019}
Dobie D.,  et~al., 2019, \mn@doi [The Astrophysical Journal]
  {10.3847/2041-8213/ab59db}, 887, L13

\bibitem[\protect\citeauthoryear{Feindt, Nordin, Rigault, Brinnel, Dhawan,
  Goobar  \& Kowalski}{Feindt et~al.}{2019}]{Feindt_2019}
Feindt U.,  Nordin J.,  Rigault M.,  Brinnel V.,  Dhawan S.,  Goobar A.,
  Kowalski M.,  2019, \mn@doi [Journal of Cosmology and Astroparticle Physics]
  {10.1088/1475-7516/2019/10/005}, 2019, 005–005

\bibitem[\protect\citeauthoryear{{Foley} \& {Mandel}}{{Foley} \&
  {Mandel}}{2013}]{2013ApJ...778..167F}
{Foley} R.~J.,  {Mandel} K.,  2013, \mn@doi [\apj]
  {10.1088/0004-637X/778/2/167}, \href
  {https://ui.adsabs.harvard.edu/abs/2013ApJ...778..167F} {778, 167}

\bibitem[\protect\citeauthoryear{Gagliano, Narayan, Engel  \& Kind}{Gagliano
  et~al.}{2021}]{Gagliano_2021}
Gagliano A.,  Narayan G.,  Engel A.,   Kind M.~C.,  2021, \mn@doi [The
  Astrophysical Journal] {10.3847/1538-4357/abd02b}, 908, 170

\bibitem[\protect\citeauthoryear{{Ghosh}, {Liu}, {Creighton}, {Kastaun},
  {Pratten}  \& {Magana Hernandez}}{{Ghosh} et~al.}{2021}]{2021arXiv210408681G}
{Ghosh} S.,  {Liu} X.,  {Creighton} J.,  {Kastaun} W.,  {Pratten} G.,   {Magana
  Hernandez} I.,  2021, arXiv e-prints, \href
  {https://ui.adsabs.harvard.edu/abs/2021arXiv210408681G} {p. arXiv:2104.08681}

\bibitem[\protect\citeauthoryear{{Goldstein} et~al.,}{{Goldstein}
  et~al.}{2017}]{2017ApJ...848L..14G}
{Goldstein} A.,  et~al., 2017, \mn@doi [\apjl] {10.3847/2041-8213/aa8f41},
  \href {https://ui.adsabs.harvard.edu/abs/2017ApJ...848L..14G} {848, L14}

\bibitem[\protect\citeauthoryear{Goldstein, Andreoni, Hankins
  et~al.}{Goldstein et~al.}{2019}]{GCN25393}
Goldstein D.,  Andreoni I.,  Hankins M.,   et~al., 2019, GCN, 25393

\bibitem[\protect\citeauthoryear{Guy et~al.,}{Guy et~al.}{2007}]{Guy_2007}
Guy J.,  et~al., 2007, \mn@doi [Astronomy \& Astrophysics]
  {10.1051/0004-6361:20066930}, 466, 11–21

\bibitem[\protect\citeauthoryear{{He}, {Zhang}, {Ren}  \& {Sun}}{{He}
  et~al.}{2015}]{2015arXiv151203385H}
{He} K.,  {Zhang} X.,  {Ren} S.,   {Sun} J.,  2015, arXiv e-prints, \href
  {https://ui.adsabs.harvard.edu/abs/2015arXiv151203385H} {p. arXiv:1512.03385}

\bibitem[\protect\citeauthoryear{Herner, Palmese, Soares-Santos  et~al.}{Herner
  et~al.}{2019}]{GCN25398}
Herner K.,  Palmese A.,  Soares-Santos M.,   et~al., 2019, GCN, 25398

\bibitem[\protect\citeauthoryear{Hlozek et~al.,}{Hlozek
  et~al.}{2020}]{plasticcresults}
Hlozek R.,  et~al., 2020, Results of the Photometric LSST Astronomical
  Time-series Classification Challenge (PLAsTiCC) (\mn@eprint {arXiv}
  {2012.12392})

\bibitem[\protect\citeauthoryear{Ho et~al.,}{Ho et~al.}{2018}]{Ho_2018}
Ho A. Y.~Q.,  et~al., 2018, \mn@doi [The Astrophysical Journal]
  {10.3847/2041-8213/aaaa62}, 854, L13

\bibitem[\protect\citeauthoryear{{Holz} \& {Hughes}}{{Holz} \&
  {Hughes}}{2005}]{holz_hughes_2005}
{Holz} D.~E.,  {Hughes} S.~A.,  2005, \mn@doi [\apj] {10.1086/431341}, \href
  {https://ui.adsabs.harvard.edu/abs/2005ApJ...629...15H} {629, 15}

\bibitem[\protect\citeauthoryear{Ivezic et~al.,}{Ivezic
  et~al.}{2019}]{Ivezi_2019}
Ivezic Z.,  et~al., 2019, \mn@doi [The Astrophysical Journal]
  {10.3847/1538-4357/ab042c}, 873, 111

\bibitem[\protect\citeauthoryear{Kasdin et~al.,}{Kasdin
  et~al.}{2020}]{Kasdin_2020}
Kasdin N.~J.,  et~al., 2020, \mn@doi [Space Telescopes and Instrumentation
  2020: Optical, Infrared, and Millimeter Wave] {10.1117/12.2562997}

\bibitem[\protect\citeauthoryear{Kasen, Metzger, Barnes, Quataert  \&
  Ramirez-Ruiz}{Kasen et~al.}{2017}]{Kasen_2017}
Kasen D.,  Metzger B.,  Barnes J.,  Quataert E.,   Ramirez-Ruiz E.,  2017,
  \mn@doi [Nature] {10.1038/nature24453}, 551, 80–84

\bibitem[\protect\citeauthoryear{{Kasliwal} et~al.,}{{Kasliwal}
  et~al.}{2017}]{kasliwal_2017}
{Kasliwal} M.~M.,  et~al., 2017, \mn@doi [Science] {10.1126/science.aap9455},
  \href {https://ui.adsabs.harvard.edu/abs/2017Sci...358.1559K} {358, 1559}

\bibitem[\protect\citeauthoryear{{Kessler} et~al.,}{{Kessler}
  et~al.}{2009}]{snana}
{Kessler} R.,  et~al., 2009, \mn@doi [\pasp] {10.1086/605984}, \href
  {https://ui.adsabs.harvard.edu/abs/2009PASP..121.1028K} {121, 1028}

\bibitem[\protect\citeauthoryear{Kessler, Conley, Jha  \& Kuhlmann}{Kessler
  et~al.}{2010a}]{kessler2010supernova}
Kessler R.,  Conley A.,  Jha S.,   Kuhlmann S.,  2010a, Supernova Photometric
  Classification Challenge (\mn@eprint {arXiv} {1001.5210})

\bibitem[\protect\citeauthoryear{{Kessler} et~al.,}{{Kessler}
  et~al.}{2010b}]{2010ApJ...717...40K}
{Kessler} R.,  et~al., 2010b, \mn@doi [\apj] {10.1088/0004-637X/717/1/40},
  \href {https://ui.adsabs.harvard.edu/abs/2010ApJ...717...40K} {717, 40}

\bibitem[\protect\citeauthoryear{{Kessler} et~al.,}{{Kessler}
  et~al.}{2019}]{plasticc}
{Kessler} R.,  et~al., 2019, \mn@doi [\pasp] {10.1088/1538-3873/ab26f1}, \href
  {https://ui.adsabs.harvard.edu/abs/2019PASP..131i4501K} {131, 094501}

\bibitem[\protect\citeauthoryear{{LIGO Scientific Collaboration, Virgo
  Collaboration}}{{LIGO Scientific Collaboration, Virgo
  Collaboration}}{2019}]{GCN25333}
{LIGO Scientific Collaboration, Virgo Collaboration} 2019, GCN, 25333

\bibitem[\protect\citeauthoryear{{Lattimer} \& {Schramm}}{{Lattimer} \&
  {Schramm}}{1974}]{lattimer_1974}
{Lattimer} J.~M.,  {Schramm} D.~N.,  1974, \mn@doi [\apjl] {10.1086/181612},
  \href {https://ui.adsabs.harvard.edu/abs/1974ApJ...192L.145L} {192, L145}

\bibitem[\protect\citeauthoryear{{Lattimer} \& {Schramm}}{{Lattimer} \&
  {Schramm}}{1976}]{1976ApJ...210..549L}
{Lattimer} J.~M.,  {Schramm} D.~N.,  1976, \mn@doi [\apj] {10.1086/154860},
  \href {https://ui.adsabs.harvard.edu/abs/1976ApJ...210..549L} {210, 549}

\bibitem[\protect\citeauthoryear{{Li} \& {Paczy{\'n}ski}}{{Li} \&
  {Paczy{\'n}ski}}{1998}]{1998ApJ...507L..59L}
{Li} L.-X.,  {Paczy{\'n}ski} B.,  1998, \mn@doi [\apjl] {10.1086/311680}, \href
  {https://ui.adsabs.harvard.edu/abs/1998ApJ...507L..59L} {507, L59}

\bibitem[\protect\citeauthoryear{Lipunov et~al.,}{Lipunov
  et~al.}{2017}]{Lipunov_2017}
Lipunov V.~M.,  et~al., 2017, \mn@doi [The Astrophysical Journal]
  {10.3847/2041-8213/aa92c0}, 850, L1

\bibitem[\protect\citeauthoryear{{Lochner}, {McEwen}, {Peiris}, {Lahav}  \&
  {Winter}}{{Lochner} et~al.}{2016}]{2016ApJS..225...31L}
{Lochner} M.,  {McEwen} J.~D.,  {Peiris} H.~V.,  {Lahav} O.,   {Winter} M.~K.,
  2016, \mn@doi [\apjs] {10.3847/0067-0049/225/2/31}, \href
  {https://ui.adsabs.harvard.edu/abs/2016ApJS..225...31L} {225, 31}

\bibitem[\protect\citeauthoryear{{Mart{\'\i}nez-Palomera}, {Bloom}  \&
  {Abrahams}}{{Mart{\'\i}nez-Palomera} et~al.}{2020}]{2020arXiv200507773M}
{Mart{\'\i}nez-Palomera} J.,  {Bloom} J.~S.,   {Abrahams} E.~S.,  2020, arXiv
  e-prints, \href {https://ui.adsabs.harvard.edu/abs/2020arXiv200507773M} {p.
  arXiv:2005.07773}

\bibitem[\protect\citeauthoryear{{Matheson} et~al.,}{{Matheson}
  et~al.}{2021}]{M21}
{Matheson} T.,  et~al., 2021, \mn@doi [\aj] {10.3847/1538-3881/abd703}, \href
  {https://ui.adsabs.harvard.edu/abs/2021AJ....161..107M} {161, 107}

\bibitem[\protect\citeauthoryear{{Metzger} et~al.,}{{Metzger}
  et~al.}{2010}]{2010MNRAS.406.2650M}
{Metzger} B.~D.,  et~al., 2010, \mn@doi [\mnras]
  {10.1111/j.1365-2966.2010.16864.x}, \href
  {https://ui.adsabs.harvard.edu/abs/2010MNRAS.406.2650M} {406, 2650}

\bibitem[\protect\citeauthoryear{{Miller} et~al.,}{{Miller}
  et~al.}{2019}]{2019ApJ...887L..24M}
{Miller} M.~C.,  et~al., 2019, \mn@doi [\apjl] {10.3847/2041-8213/ab50c5},
  \href {https://ui.adsabs.harvard.edu/abs/2019ApJ...887L..24M} {887, L24}

\bibitem[\protect\citeauthoryear{Morgan et~al.,}{Morgan
  et~al.}{2020}]{Morgan_2020}
Morgan R.,  et~al., 2020, \mn@doi [The Astrophysical Journal]
  {10.3847/1538-4357/abafaa}, 901, 83

\bibitem[\protect\citeauthoryear{{Muthukrishna}, {Narayan}, {Mandel}, {Biswas}
  \& {Hlo{\v{z}}ek}}{{Muthukrishna} et~al.}{2019}]{rapid_2019}
{Muthukrishna} D.,  {Narayan} G.,  {Mandel} K.~S.,  {Biswas} R.,
  {Hlo{\v{z}}ek} R.,  2019, \mn@doi [\pasp] {10.1088/1538-3873/ab1609}, \href
  {https://ui.adsabs.harvard.edu/abs/2019PASP..131k8002M} {131, 118002}

\bibitem[\protect\citeauthoryear{{Narayan} et~al.,}{{Narayan}
  et~al.}{2018a}]{2018ApJS..236....9N}
{Narayan} G.,  et~al., 2018a, \mn@doi [\apjs] {10.3847/1538-4365/aab781}, \href
  {https://ui.adsabs.harvard.edu/abs/2018ApJS..236....9N} {236, 9}

\bibitem[\protect\citeauthoryear{{Narayan} et~al.,}{{Narayan}
  et~al.}{2018b}]{N18}
{Narayan} G.,  et~al., 2018b, \mn@doi [\apjs] {10.3847/1538-4365/aab781}, \href
  {https://ui.adsabs.harvard.edu/abs/2018ApJS..236....9N} {236, 9}

\bibitem[\protect\citeauthoryear{Radice, Perego, Hotokezaka, Fromm, Bernuzzi
  \& Roberts}{Radice et~al.}{2018}]{Radice_2018}
Radice D.,  Perego A.,  Hotokezaka K.,  Fromm S.~A.,  Bernuzzi S.,   Roberts
  L.~F.,  2018, \mn@doi [The Astrophysical Journal] {10.3847/1538-4357/aaf054},
  869, 130

\bibitem[\protect\citeauthoryear{Remy}{Remy}{2020}]{KerasTCN}
Remy P.,  2020, Temporal Convolutional Networks for Keras,
  \url{https://github.com/philipperemy/keras-tcn}

\bibitem[\protect\citeauthoryear{{Schutz}}{{Schutz}}{1986}]{schutz_1986}
{Schutz} B.~F.,  1986, \mn@doi [\nat] {10.1038/323310a0}, \href
  {https://ui.adsabs.harvard.edu/abs/1986Natur.323..310S} {323, 310}

\bibitem[\protect\citeauthoryear{Shibata \& Hotokezaka}{Shibata \&
  Hotokezaka}{2019}]{Shibata_2019}
Shibata M.,  Hotokezaka K.,  2019, \mn@doi [Annual Review of Nuclear and
  Particle Science] {10.1146/annurev-nucl-101918-023625}, 69, 41–64

\bibitem[\protect\citeauthoryear{Singer \& Price}{Singer \&
  Price}{2016}]{singer_and_price}
Singer L.~P.,  Price L.~R.,  2016, \mn@doi [Phys. Rev. D]
  {10.1103/PhysRevD.93.024013}, 93, 024013

\bibitem[\protect\citeauthoryear{Smartt, Smith, Srivastav  et~al.}{Smartt
  et~al.}{2019}]{GCN25455}
Smartt S.,  Smith K.~W.,  Srivastav S.,   et~al., 2019, GCN, 25455

\bibitem[\protect\citeauthoryear{Soares-Santos et~al.,}{Soares-Santos
  et~al.}{2017}]{Soares_Santos_2017}
Soares-Santos M.,  et~al., 2017, \mn@doi [The Astrophysical Journal]
  {10.3847/2041-8213/aa9059}, 848, L16

\bibitem[\protect\citeauthoryear{Stachie, Coughlin, Christensen  \&
  Muthukrishna}{Stachie et~al.}{2020}]{Stachie_2020}
Stachie C.,  Coughlin M.~W.,  Christensen N.,   Muthukrishna D.,  2020, \mn@doi
  [Monthly Notices of the Royal Astronomical Society] {10.1093/mnras/staa1776},
  497, 1320–1331

\bibitem[\protect\citeauthoryear{{Street}, {Bowman}, {Saunders}  \&
  {Boroson}}{{Street} et~al.}{2018}]{TOMS}
{Street} R.~A.,  {Bowman} M.,  {Saunders} E.~S.,   {Boroson} T.,  2018, in
  {Guzman} J.~C.,  {Ibsen} J.,  eds,  Society of Photo-Optical Instrumentation
  Engineers (SPIE) Conference Series Vol. 10707, Software and
  Cyberinfrastructure for Astronomy V. p. 1070711 (\mn@eprint {arXiv}
  {1806.09557}), \mn@doi{10.1117/12.2312293}

\bibitem[\protect\citeauthoryear{{Street} et~al.,}{{Street}
  et~al.}{2020}]{AEON}
{Street} R.~A.,  et~al., 2020, in Society of Photo-Optical Instrumentation
  Engineers (SPIE) Conference Series. p. 1144925, \mn@doi{10.1117/12.2559986}

\bibitem[\protect\citeauthoryear{T.-W.~Chen et~al.}{T.-W.~Chen
  et~al.}{2019}]{GCN25457}
T.-W.~Chen T.~Schweyer A. N.~G.,  et~al., 2019, GCN, 25457

\bibitem[\protect\citeauthoryear{Tanvir, Levan, Fruchter, Hjorth, Hounsell,
  Wiersema  \& Tunnicliffe}{Tanvir et~al.}{2013}]{Tanvir_2013}
Tanvir N.~R.,  Levan A.~J.,  Fruchter A.~S.,  Hjorth J.,  Hounsell R.~A.,
  Wiersema K.,   Tunnicliffe R.~L.,  2013, \mn@doi [Nature]
  {10.1038/nature12505}, 500, 547–549

\bibitem[\protect\citeauthoryear{Tanvir et~al.,}{Tanvir
  et~al.}{2017}]{Tanvir_2017}
Tanvir N.~R.,  et~al., 2017, \mn@doi [The Astrophysical Journal]
  {10.3847/2041-8213/aa90b6}, 848, L27

\bibitem[\protect\citeauthoryear{Tohuvavohu, Kennea, DeLaunay, Palmer, Cenko
  \& Barthelmy}{Tohuvavohu et~al.}{2020}]{Tohuvavohu_2020}
Tohuvavohu A.,  Kennea J.~A.,  DeLaunay J.,  Palmer D.~M.,  Cenko S.~B.,
  Barthelmy S.,  2020, \mn@doi [The Astrophysical Journal]
  {10.3847/1538-4357/aba94f}, 900, 35

\bibitem[\protect\citeauthoryear{V.~Lipunov et~al.}{V.~Lipunov
  et~al.}{2019}]{GCN25474}
V.~Lipunov D.~Vlasenko E.~G.,  et~al., 2019, GCN, 25474

\bibitem[\protect\citeauthoryear{Valenti et~al.,}{Valenti
  et~al.}{2017}]{Valenti_2017}
Valenti S.,  et~al., 2017, \mn@doi [The Astrophysical Journal]
  {10.3847/2041-8213/aa8edf}, 848, L24

\bibitem[\protect\citeauthoryear{Vieira et~al.,}{Vieira
  et~al.}{2020}]{Vieira_2020}
Vieira N.,  et~al., 2020, \mn@doi [The Astrophysical Journal]
  {10.3847/1538-4357/ab917d}, 895, 96

\bibitem[\protect\citeauthoryear{Vincenzi, Sullivan, Firth, Gutiérrez,
  Frohmaier, Smith, Angus  \& Nichol}{Vincenzi et~al.}{2019}]{Vincenzi_2019}
Vincenzi M.,  Sullivan M.,  Firth R.~E.,  Gutiérrez C.~P.,  Frohmaier C.,
  Smith M.,  Angus C.,   Nichol R.~C.,  2019, \mn@doi [Monthly Notices of the
  Royal Astronomical Society] {10.1093/mnras/stz2448}, 489, 5802–5821

\bibitem[\protect\citeauthoryear{Watson et~al.,}{Watson
  et~al.}{2020}]{Watson_2020}
Watson A.~M.,  et~al., 2020, \mn@doi [Monthly Notices of the Royal Astronomical
  Society] {10.1093/mnras/staa161}, 492, 5916–5921

\bibitem[\protect\citeauthoryear{X.~Wang et~al.}{X.~Wang
  et~al.}{2019}]{GCN25485}
X.~Wang S.~Antier M.~C.,  et~al., 2019, GCN, 25485

\bibitem[\protect\citeauthoryear{de Wet et~al.,}{de~Wet
  et~al.}{2021}]{de_Wet_2021}
de Wet S.,  et~al., 2021, \mn@doi [Astronomy \& Astrophysics]
  {10.1051/0004-6361/202040231}, 649, A72

\makeatother
\end{thebibliography}

\bsp	% typesetting comment
\label{lastpage}
\end{document}